\pdfoutput=1

\documentclass[twocolumn]{emulateapj}

\usepackage{amsmath}
\usepackage{graphicx}
\usepackage{color}
\usepackage[colorlinks]{hyperref}
\hypersetup{
    colorlinks,	
    citecolor=blue,
}

\newcommand{\be}{\begin{eqnarray}}
\newcommand{\ee}{\end{eqnarray}}

\def\jcap{JCAP}

\shorttitle{Testing the Kerr nature of GRS~1716--249}
\shortauthors{Zhang et al.}

\begin{document}

\title{Testing the Kerr black hole hypothesis with GRS~1716--249\\by combining the continuum-fitting and the iron-line methods}

\author{Zuobin~Zhang\altaffilmark{1}, Honghui~Liu\altaffilmark{1}, Askar~B.~Abdikamalov\altaffilmark{1,2}, Dimitry~Ayzenberg\altaffilmark{3}, Cosimo~Bambi\altaffilmark{1,\dag}, and Menglei~Zhou\altaffilmark{4}}

\altaffiltext{1}{Center for Field Theory and Particle Physics and Department of Physics, 
Fudan University, 200438 Shanghai, China. \email[\dag E-mail: ]{bambi@fudan.edu.cn}}
\altaffiltext{2}{Ulugh Beg Astronomical Institute, Tashkent 100052, Uzbekistan}
\altaffiltext{3}{Theoretical Astrophysics, Eberhard-Karls Universit\"at T\"ubingen, D-72076 T\"ubingen, Germany}
\altaffiltext{4}{Institut f\"ur Astronomie und Astrophysik, Eberhard-Karls Universit\"at T\"ubingen, D-72076 T\"ubingen, Germany}

\begin{abstract}
The continuum-fitting and the iron-line methods are currently the two leading techniques for measuring the spins of accreting black holes. In the past few years, these two methods have been developed for testing fundamental physics. In the present work, we employ state-of-the-art models to test black holes through the continuum-fitting and the iron-line methods and we analyze three \textsl{NuSTAR} observations of the black hole binary GRS~1716--249 during its outburst in 2016-2017. In these three observations, the source was in a hard-intermediate state and the spectra show both a strong thermal component and prominent relativistic reflection features. Our analysis confirms the Kerr nature of the black hole in GRS~1716--249 and provides quite stringent constraints on possible deviations from the predictions of general relativity. 
\end{abstract}


\section{Introduction}

The theory of general relativity was proposed by Einstein at the end of 1915. Over the decades, it has successfully passed a large number of observational tests and has thus survived to now without any modifications. For many years, the theory has been mainly tested in the so-called ``weak field'' regime with experiments in the solar system and observations of binary pulsars~\citep{2014LRR....17....4W}. The situation has significantly changed in the past five years, and we have now a number of observational tests in the ``strong field'' regime with gravitational waves~\citep{2016PhRvL.116v1101A,2019PhRvD.100j4036A,2016PhRvD..94h4002Y}, X-ray observations~\citep{2018PhRvL.120e1101C,2019ApJ...875...56T,2021ApJ...907...31T}, and mm VLBI data~\citep{2019PhRvD.100d4057B,2019PhRvD.100b4020V,2020PhRvL.125n1104P}.

Black holes are ideal laboratories for testing gravity in the strong field regime~\citep{2017RvMP...89b5001B,2017bhlt.book.....B,2016CQGra..33e4001Y,2019LRR....22....4C}. In 4-dimensional general relativity and in the absence of exotic matter fields, uncharged black holes are described by the Kerr solution~\citep{1963PhRvL..11..237K}, and every object is completely characterized by its mass $M$ and spin angular momentum $J$~\citep[see, e.g.,][]{2012LRR....15....7C}. The spacetime metric around astrophysical black holes is thought to be approximated well by the Kerr solution, and deviations from the Kerr geometry induced by nearby stars, the presence of an accretion disk, or a non-vanishing electric charge are normally completely negligible~\cite[see, e.g.,][]{2014PhRvD..89l7302B,2017bhlt.book.....B}. However, macroscopic deviations from the Kerr metric are expected in scenarios with large quantum gravity effects at the black hole event horizon~\citep[see, e.g.,][]{2017NatAs...1E..67G}, in extensions of general relativity~\citep[see, e.g.,][]{2011PhRvL.106o1104K}, and in the presence of exotic matter fields~\citep[see, e.g.,][]{2014PhRvL.112v1101H}. Testing the Kerr metric around an astrophysical black hole is thus a method for testing new physics in the strong gravity regime.

X-ray observations can test black holes accreting from geometrically thin and optically thick disks. The gas at every point of the accretion disk is in local thermal equilibrium and emits a blackbody-like spectrum. The whole disk has a multi-temperature blackbody-like spectrum and the emission is peaked in the soft X-ray band for stellar-mass black holes and in the UV band for supermassive black holes. The so-called ``corona'' is some hotter ($\sim 100$~keV) gas near the black hole; it may be the accretion flow between the inner edge of the disk and the black hole, the atmosphere above the accretion disk, the base of the jet, etc. Thermal photons from the accretion disk can inverse Compton scatter off free electrons in the corona. The Comptonized photons can illuminate the disk: Compton scattering and absorption followed by fluorescent emission generate a reflection component. For more details, see, for instance, \citet{2021SSRv..217...65B} and references therein.

The continuum-fitting method is the analysis of the thermal spectrum of thin accretion disks~\citep{1997ApJ...482L.155Z,2006ApJ...652..518M,2014SSRv..183..295M}. X-ray reflection spectroscopy (or iron-line method) denotes the analysis of the reflection features of thin accretion disks~\citep{1989MNRAS.238..729F,2006ApJ...652.1028B,2021SSRv..217...65B}. Both methods were originally proposed and developed for measuring black hole spins assuming standard physics and have been later extended for testing new physics~\citep[preliminary studies to use these techniques for testing new physics were reported in][]{2002NuPhB.626..377T,2003IJMPD..12...63L,2008PhRvD..78b4043P,2009GReGr..41.1795S,2011ApJ...731..121B,2013ApJ...773...57J,2013PhRvD..87b3007B}.

Within the standard framework, the calculations of the thermal spectrum and the reflection features of the disk assume that the background metric is described by the Kerr solution, that all particles follow the geodesics of the spacetime, and -- for the calculations of the reflection features -- the atomic physics in the strong gravitational field of the black hole is the same as in our laboratories on Earth. Relaxing these assumptions it is possible to test the Kerr metric, the geodesic motion of particles, and the atomic physics in the strong gravitational field of black holes. There are indeed a number of scenarios of new physics where these assumptions can be violated.

The state-of-the-art in the field is represented by two recent XSPEC models, called {\tt nkbb}~\citep{2019PhRvD..99j4031Z} and {\tt relxill\_nk}~\citep{2017ApJ...842...76B,2019ApJ...878...91A,2020ApJ...899...80A}. {\tt nkbb} is to analyze the thermal spectrum of an accretion disk and {\tt relxill\_nk} is for studying its reflection features. In the past few years, these models have been employed to analyze a number of black hole X-ray data and testing fundamental physics. As of now, most efforts have been devoted to testing the Kerr metric around both stellar-mass and supermassive black holes \citep[see, however,][where {\tt relxill\_nk} is used to test the Weak Equivalence Principle in the strong gravitational field of the black hole in EXO~1846--031]{2021PhRvD.104d4001R}. Most of the past studies have employed the parametric black hole spacetime proposed by Johannsen~\citep{2013PhRvD..88d4002J} to constrain its deformation parameter $\alpha_{13}$. This is simply because, for the moment, efforts have been devoted to improving the model and understanding the systematic uncertainties, while less importance has been given to test specific scenarios of new physics. However, {\tt nkbb} and {\tt relxill\_nk} can be easily modified to test any black hole solution known in analytic form~\citep[see, e.g.,][]{2020EPJC...80..622Z,2021JCAP...01..047Z,2021JCAP...07..002T}.

In the present work, we analyze three \textsl{NuSTAR} observations with simultaneous or quasi-simultaneous \textsl{Swift} snapshots of the black hole binary GRS~1716--249 during its outburst in 2016-2017. In these three observations, the source was in a hard-intermediate state and the three spectra show a strong thermal spectrum and prominent relativistic reflection features, which allows us to simultaneously use {\tt nkbb} and {\tt relxill\_nk} in our spectral analysis. A similar situation was encountered in the analysis of the \textsl{NuSTAR} observation of GX~339--4 reported in \citet{2021ApJ...907...31T}, but in that case {\tt nkbb} was used to estimate the black hole mass and distance, which are not known from optical observations in the case of GX~339--4, and therefore we cannot speak of the continuum-fitting method. In the case of GRS~1716--249, we have an independent estimate of the black hole mass and distance, which can thus help to constrain possible deviations from the Kerr geometry.

The content of the paper is as follows. In Section~\ref{s-red}, we present the observations analyzed in our study and we describe our data reduction. In Section~\ref{s-sa}, we present the spectral analysis of every observation separately and of all observations together. We discuss our results in Section~\ref{s-dis}. The Johannsen metric employed in our model is briefly reviewed in the appendix.


\section{Observations and data reduction}\label{s-red}

\subsection{Observations}

\begin{figure}[t]
	\centering
	\includegraphics[scale=0.55]{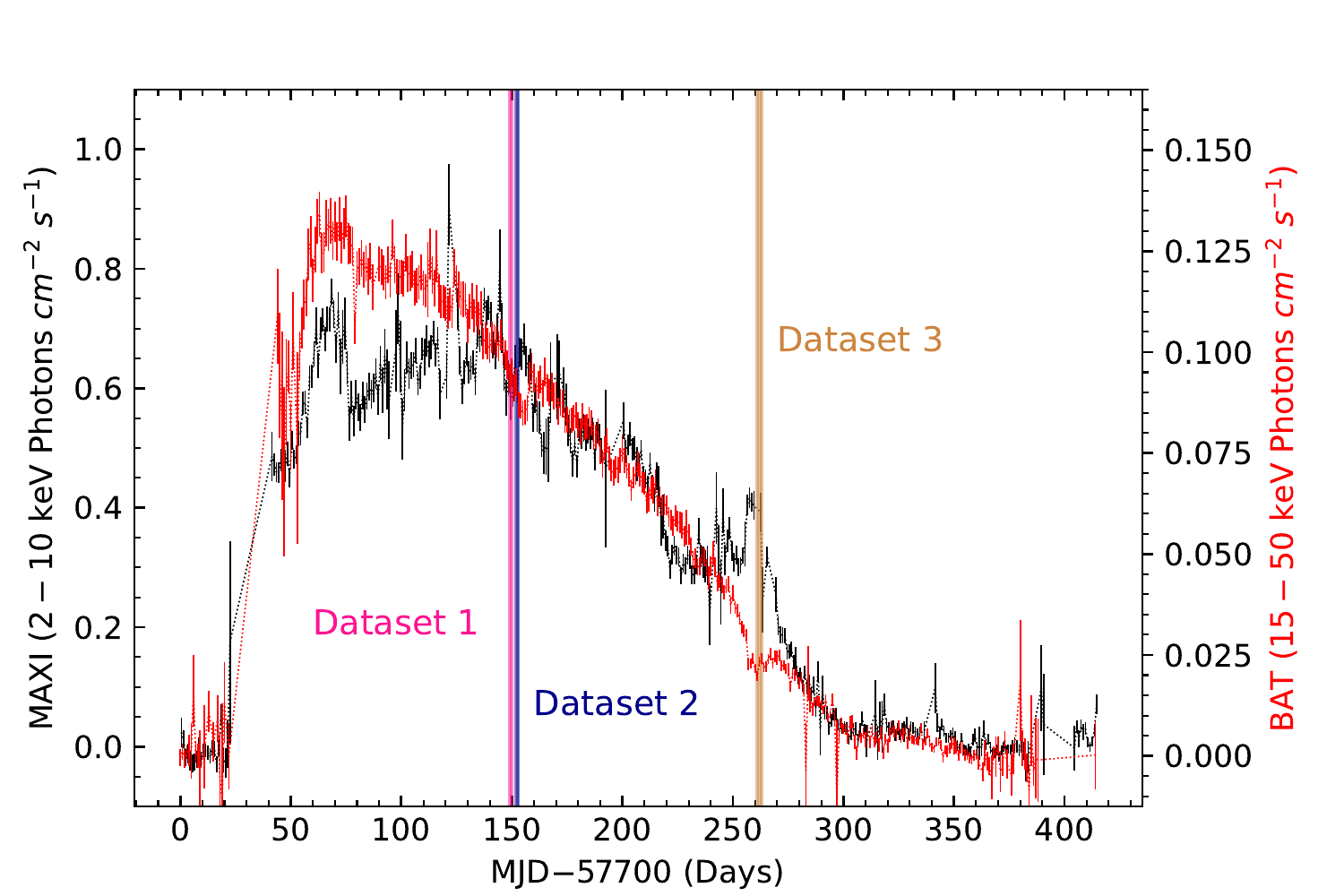}\\
	\vspace{0.2cm}
	\caption{Light curves of GRS~1716--249 during its 2016-2017 outburst from \textsl{Swift}/BAT (15-50~keV, red data) and \textsl{MAXI} (2-10~keV, black data). The pink, blue, and brown bars represent the time of the \textsl{Swift} and \textsl{NuSTAR} observations analyzed in the present work.} 
	\label{f-c1}	 
\end{figure}

The Galactic black hole binary GRS~1716--249, also known as GRO~J1719--24, was discovered in September 1993 by \textsl{CGRO}/BATSE and \textsl{Granat}/SIGMA \citep{1994IAUC.6104....1H,1993IAUC.5874....1B}. After more than twenty years in quiescence, GRS~1716--249 exhibited an outburst in 2016-2017. This outburst lasted about nine months and the source was observed by four X-ray missions: \textsl{NuSTAR}, \textsl{Swift}, \textsl{MAXI}, and \textsl{AstroSat}. Fig.~\ref{f-c1} shows the long-term light curves in the 2-10~keV and 15-20~keV bands observed with \textsl{MAXI} (in black) and \textsl{Swift}/BAT (in red), respectively. According to the classification method used by \citet{2010A&A...520A..98D}, there are two types of outbursts: \emph{fast rise slow decay} (FRSD) and \emph{slow rise slow decay} (SRSD), depending on the variation of flux or count rates in light curve profiles. From the \textsl{MAXI} and \textsl{Swift}/BAT light curve profiles shown in Fig.~\ref{f-c1}, we notice that both fluxes increase significantly in a very short time and start decreasing slowly, so we have a typical FRSD outburst.

In the 2016-2017 outburst, the source approached the high/soft state three times but never reached a true high/soft state~\citep{2019MNRAS.482.1587B}. In this paper, we analyze three \textsl{NuSTAR} observations with simultaneous/quasi-simultaneous \textsl{Swift} observations in which the source was closer to a high/soft state. These three observations are marked by the pink, blue, and brown bars in Fig.~\ref{f-c1}. Table~\ref{t-obs} shows the details of these three observations.


\begin{table*}
\centering
\vspace{0.5cm}
\begin{tabular}{ccccccc}
\hline\hline
Dataset &  \hspace{0.5cm} Mission \hspace{0.5cm}  & \hspace{0.5cm} Obs.~ID \hspace{0.5cm} & \hspace{0.5cm} Instrument \hspace{0.5cm} & \hspace{0.5cm} Start date \hspace{0.5cm} & \hspace{0.5cm} Exposure~(ks) \hspace{0.5cm} \\
\hline\hline
1 & \textsl{NuSTAR} & 90202055002 & FPMA+FPMB & 2017-04-07 & 17.9 \\
  & \textsl{Swift} & 00034924029 & XRT & 2017-04-07 & 1.7 \\
  & \textsl{Swift} & 00034924030 & XRT & 2017-04-08 & 1.0 \\
\hline
2 & \textsl{NuSTAR} & 90202055004 & FPMA+FPMB & 2017-04-10 & 15.8 \\
   & \textsl{Swift} & 00034924031 & XRT & 2017-04-10 & 1.9 \\
\hline
3 & \textsl{NuSTAR} & 90301007002 & FPMA+FPMB & 2017-07-28 & 89.3 \\
   & \textsl{Swift} & 00034924051 & XRT & 2017-07-29 & 1.0 \\
   & \textsl{Swift} & 00034924052 & XRT & 2017-07-30 & 1.0 \\
\hline\hline
\end{tabular}
\vspace{0.3cm}
\caption{\rm Summary of the observations analyzed in the present work. \label{t-obs}}
\end{table*}


\subsection{Data reduction}

We reduce the \textsl{NuSTAR} data using the standard pipeline NUPIPELINE v0.4.6 in HEASOFT v6.26. The calibration file version is v20210315. For two observations (Obs.~ID 90202055002 and 90202055004), we create their clean event files by setting the keyword {\tt statusexpr} to be “STATUS==b0000xxx00xxxx000”, following \citet{2019ApJ...887..184T} and the \textsl{NuSTAR} analysis guide. The source events are extracted from a circular region centered on GRS~1716--249 with a radius of 200~arcsec. The background region is a circle with the same size taken far from the source region to avoid any contribution from the source. We use the FTOOL NUPRODUCTS to generate the spectra and other products. The FPMA and FPMB spectra are grouped to have a minimum count of 50~photons per bin. The \textsl{NuSTAR} data are modeled over the full 3-78~keV band in this work.

In order to extend these spectra below 3~keV, we try to include simultaneous \textsl{Swift} data in our analysis. For one of the \textsl{NuSTAR} observations (Obs.~ID 90202055002), there are no strict simultaneous \textsl{Swift} observations. We thus use two quasi-simultaneous observations instead, whose time differences are less than 20 hours~\citep{2019ApJ...887..184T}.

We use the standard pipeline XRTPIPELINE v0.13.5 in HEASOFT v6.26 to reduce the \textsl{Swift} data and we only include the grade 0 events~\citep{2019ApJ...887..184T}. The calibration file version is v20200724. An issue in the \textsl{Swift} data of GRS~1716--249 is the presence of photon pile-up, as a consequence of the brightness of the source. To minimize the pile-up effect, we exclude the most heavily piled-up region~\citep{2019MNRAS.482.1587B,2019ApJ...887..184T}. Finally, we extract the source events from an annulus region, with an inner radius of 3~pixels and outer radius of 40~pixels, which is centered on the source coordinates. The background events are extracted from an annulus region of the same size far from the source region. We generate the spectra and other products with FTOOL XRTPRODUCTS. The spectra are grouped to have a minimum count of 50~photons per bin and are fit over the 0.8-8.0~keV band.


\section{Spectral analysis}\label{s-sa}

\begin{figure}[t]
	\centering
	\includegraphics[scale=0.5]{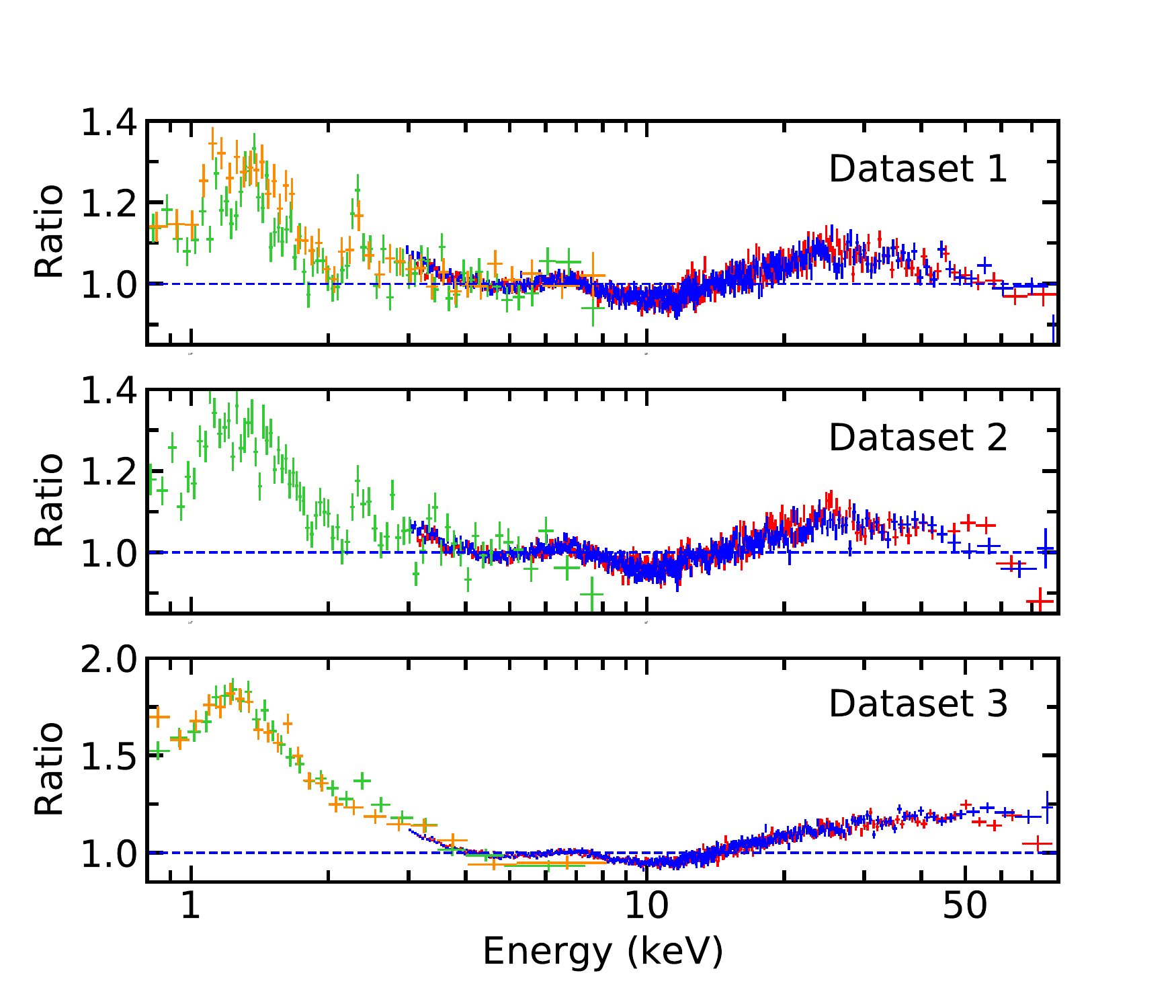} \\
	\caption{Ratio plots of \textsl{NuSTAR}/FPMA (red), \textsl{NuSTAR}/FPMB (blue), and \textsl{Swift}/XRT (green and orange) when we fit the data with an absorbed power-law spectrum. In the fit, we ignore the data below 3~keV and above 78~keV. All datasets show a broad iron line and a strong Compton hump. The data are rebinned for visual clarity.} 
	\label{ratio_1}	 
\end{figure}

We employ XSPEC v12.10.1s for the spectral analysis. First, we fit the three spectra separately with an absorbed power-law model, i.e. {\tt tbabs$\times$cutoffpl} in XSPEC language. {\tt tbabs} accounts for the Galactic column absorption~\citep{2000ApJ...542..914W} and {\tt cutoffpl} for a power-law component from the corona. The resulting data to best-fit model ratios are shown in Fig.~\ref{ratio_1}. We clearly see a small feature around 6-7~keV, and a bump above 10~keV in all spectra, which can be naturally interpreted as an iron line and a Compton hump from the reflection spectrum. In addition to these relativistic reflection features, there is an excess of photons at low energies (below 2~keV), which suggests the presence of the thermal component from the accretion disk.


\begin{table*}[]
	\centering
	{\renewcommand{\arraystretch}{1.3}
	\begin{tabular}{lcc|cc|cc}
		\hline\hline
		Parameter & \multicolumn{2}{c}{Dataset 1} & \multicolumn{2}{c}{Dataset 2} & \multicolumn{2}{c}{Dataset 3} \\ \hline
		& $\alpha_{13}=0$ & $\alpha_{13}$ free & $\alpha_{13}=0$ & $\alpha_{13}$ free & $\alpha_{13}=0$ & $\alpha_{13}$ free \\ 
		\hline
		{\tt tbabs}   \\
		$N_{\rm H}$ [$10^{22}$~cm$^{-2}$] & $0.606_{-0.018}^{+0.015}$ & $0.606_{-0.018}^{+0.015}$ & $0.607_{-0.011}^{+0.022}$ & $0.608_{-0.018}^{+0.021}$ & $0.608_{-0.018}^{+0.020}$ & $0.610_{-0.020}^{+0.018}$ \\ \hline
		{\tt nkbb}  \\  
		$M$ [$M_{\odot}$]  & $5.0^*$ & $5.0^*$ & $5.0^*$ & $5.0^*$ & $5.0^*$ & $5.0^*$ \\
		$\dot{M}$ [$10^{15}$~g~s$^{-1}$] & $4.0_{-0.5}^{+0.3}$ & $3.96_{-0.5}^{+0.21}$ & $5.2_{-0.3}^{+2.0}$ & $5.1_{-0.4}^{+0.6}$ & $8.1_{-0.9}^{+1.0}$ & $7.9_{-0.6}^{+1.2}$ \\
		$D$ [kpc]  & $2.4^*$ & $2.4^*$ & $2.4^*$ & $2.4^*$ & $2.4^*$ & $2.4^*$ \\
		$f_{\rm col}$  & $1.7^*$ & $1.7^*$ & $1.7^*$ & $1.7^*$ & $1.7^*$ & $1.7^*$ \\ \hline
		{\tt relxill\_nk}  \\ 
		$ q_{\rm in} $& $6.2_{-0.6}^{+0.5}$ & $6.1_{-0.9}^{+0.5}$ & $4.9_{-0.9}^{+1.1}$ & $4.9_{-1.4}^{+0.9}$ & $3^*$ & $3^*$ \\ 
		$ q_{\rm out} $& $1^*$ & $1^*$ & $1^*$ & $1^*$ & $3^*$ & $3^*$ \\ 
		$ R_{\rm br}$ [$r_{\rm g}$] & $9.4_{-1.3}^{+2.7}$ & $9.7_{-2.0}^{+6}$ & $15_{-6}^{+15}$ & $16_{-8}^{+23}$ & $15^*$ & $15^*$ \\ 
		$ a_* $& $>0.992$ & $>0.990$ & $>0.981$ & $>0.967$ & $0.948_{-0.015}^{+0.016}$ & $>0.859$ \\ 
		$ i $ [deg]& $61.2_{-2.3}^{+1.6}$ & $61\pm4$& $52_{-4}^{+3}$ & $53_{-8}^{+3}$ & $48.9_{-0.9}^{+1.1}$ & $48.8\pm1.1$ \\ 
		$ \Gamma $& $1.635_{-0.013}^{+0.008}$ & $1.635_{-0.013}^{+0.008}$ & $1.691_{-0.015}^{+0.013}$ & $1.690_{-0.012}^{+0.015}$ & $1.930_{-0.016}^{+0.006}$ & $1.931_{-0.019}^{+0.006}$  \\ 
		$\log\xi$ & $3.16_{-0.10}^{+0.09}$ & $3.16_{-0.08}^{+0.09}$& $3.42_{-0.09}^{+0.07}$ & $3.40_{-0.07}^{+0.09}$ & $3.44_{-0.05}^{+0.07}$ & $3.44_{-0.04}^{+0.10}$ \\ 
		$ A_{\rm Fe} $& $5.0_{-1.3}^{+0.4}$ & $5.0_{-0.6}^{+0.5}$ & $5.0_{-0.8}^{+1.2}$ & $5.0_{-0.6}^{+1.3}$ & $3.4_{-0.3}^{+0.9}$ & $3.4_{-0.3}^{+1.1}$ \\ 
		$ E_{\rm cut} $ [keV]  &  $175_{-13}^{+39}$  & $175_{-13}^{+41}$& $196_{-29}^{+34}$ & $190_{-21}^{+39}$ & $>663$ & $>642$ \\ 
		$ R_{\rm ref} $ & $0.095_{-0.006}^{+0.025}$ & $0.097\pm0.008$& $0.097_{-0.009}^{+0.014}$ & $0.096_{-0.007}^{+0.015}$ & $0.131_{-0.017}^{+0.015}$ & $0.131_{-0.018}^{+0.013}$ \\ 
		$ \alpha_{13} $& $0^*$ & $0.00_{-0.21}^{+0.05}$ &  $0^*$ &  $0.07_{-0.5}^{+0.05}$ & $0^*$ &  $0.28_{-1.0}^{+0.03}$ \\ 
		Norm& $0.0258_{-0.0006}^{+0.0005}$ & $0.0257_{-0.0006}^{+0.0007}$ &  $0.0237_{-0.0008}^{+0.0007}$ & $0.0237_{-0.0006}^{+0.0007}$ & $0.0086\pm0.0006$ & $0.0087\pm0.0003$ \\ \hline
		Cross-normalization \\
		$C_{\rm FPMB}$ & $1.0136\pm0.0015$ & $1.0136_{-0.0022}^{+0.0015}$ & $1.0147\pm0.0016$ & $1.0147\pm0.0016$ & $1.0158\pm0.0012$ & $1.0158\pm0.0012$ \\
		$C_{\rm XRT1}$ & $0.729_{-0.007}^{+0.006}$ & $0.729\pm0.007$ & $0.712_{-0.008}^{+0.007}$ & $0.712_{-0.008}^{+0.007}$ & $0.977_{-0.024}^{+0.016}$ & $0.977_{-0.015}^{+0.016}$\\
		$C_{\rm XRT2}$ & $0.947_{-0.010}^{+0.009}$ & $0.948\pm0.010$ & & & $0.847_{-0.015}^{+0.014}$ & $0.847_{-0.013}^{+0.015}$ \\ \hline
		$\chi^2/\nu $ & $3857.39/3500$ & $3857.36/3499$ & $3122.42/2959$ & $3122.01/2958$ & $3578.78/3231$ & $3577.79/3230$ \\ 
		& $=1.10211$ & $=1.10242$ & $=1.05523$ & $=1.05545$ & $=1.10764$ & $=1.10768$ \\
		\hline\hline
	\end{tabular}
	}
	\caption{\rm Best-fit values when we fit every dataset separately. The reported uncertainties correspond to the 90\% confidence level for one relevant parameter ($\Delta\chi^2 = 2.71$). $^*$ indicates that the parameter is frozen in the fit. $\xi$ is in units of erg~cm~s$^{-1}$. See the text for more details. 
	\label{t-bestfit_1}}
\end{table*}


\begin{figure*}[t]
	\begin{center}
		\includegraphics[scale=0.45]{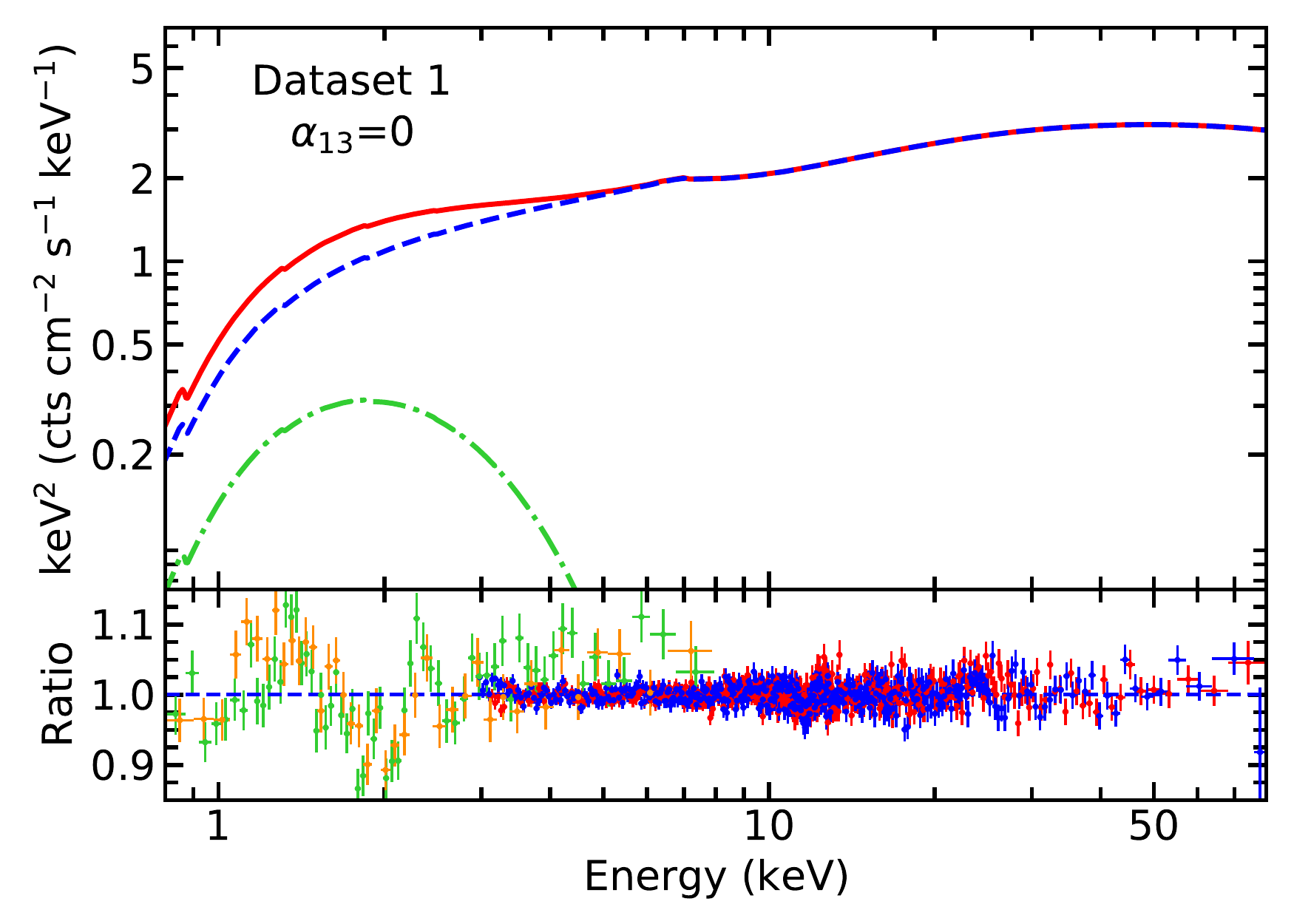}
		\hspace{0.5cm}
		\includegraphics[scale=0.45]{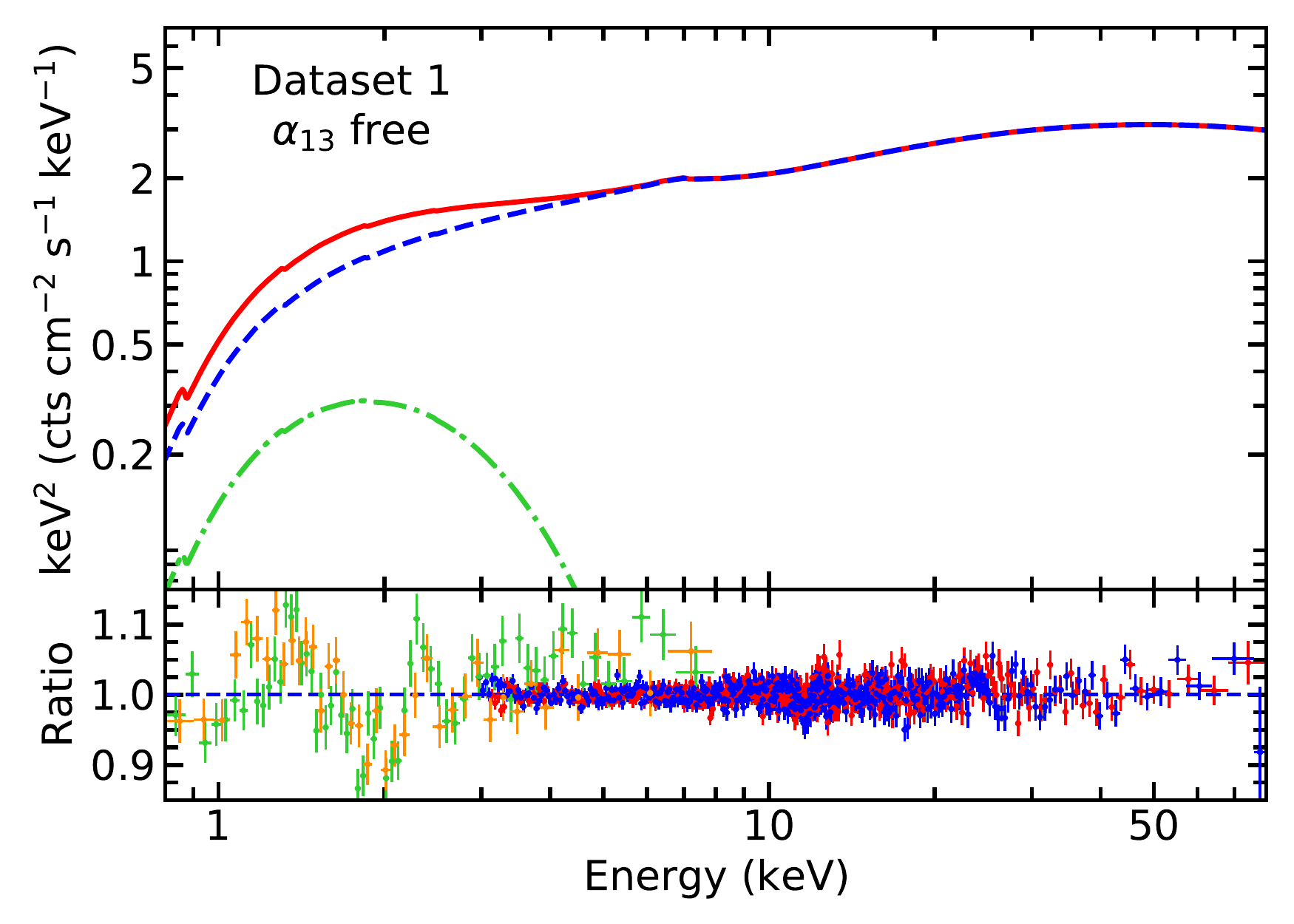} \\ \vspace{0.3cm}
		\includegraphics[scale=0.45]{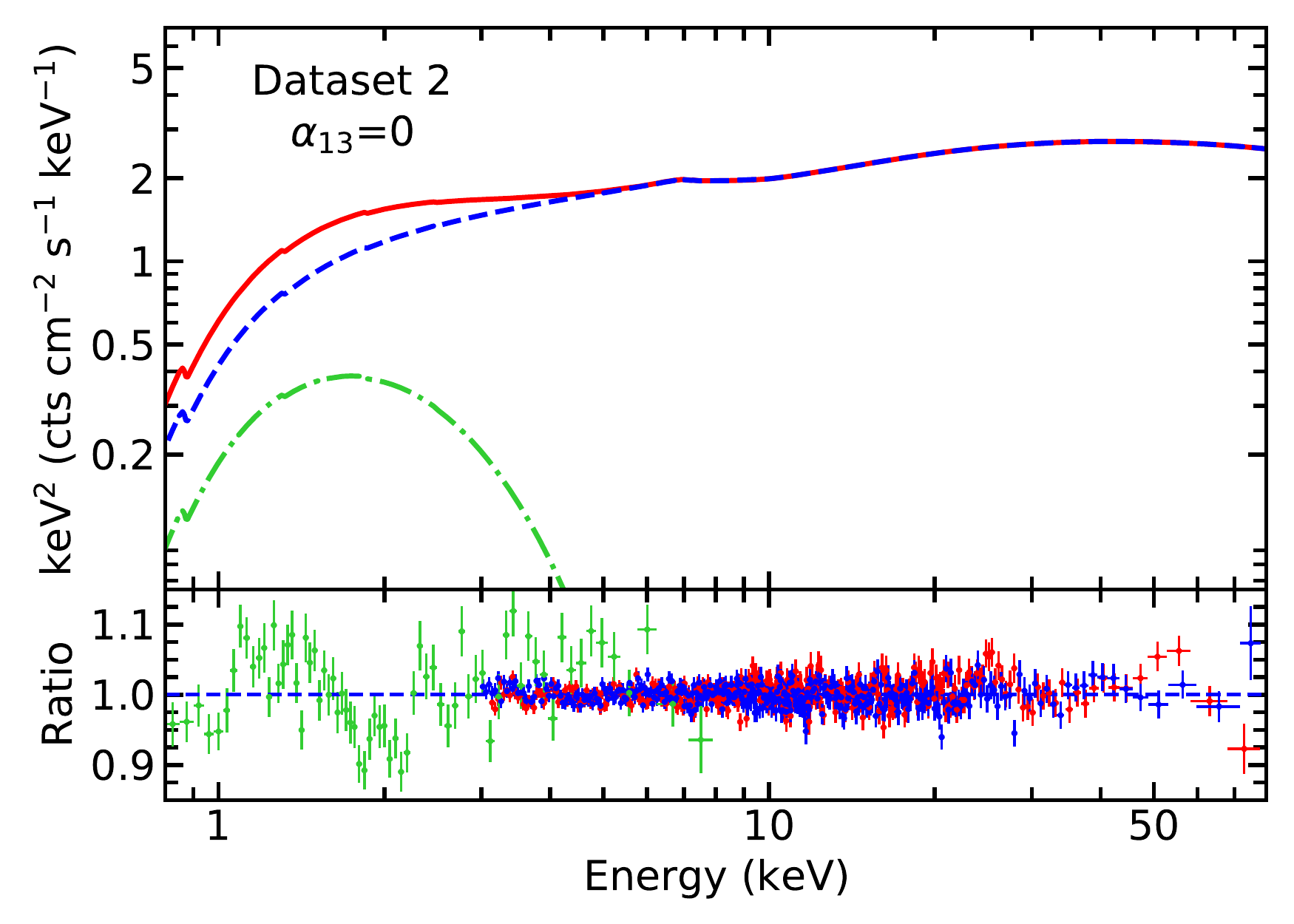}
		\hspace{0.5cm}
		\includegraphics[scale=0.45]{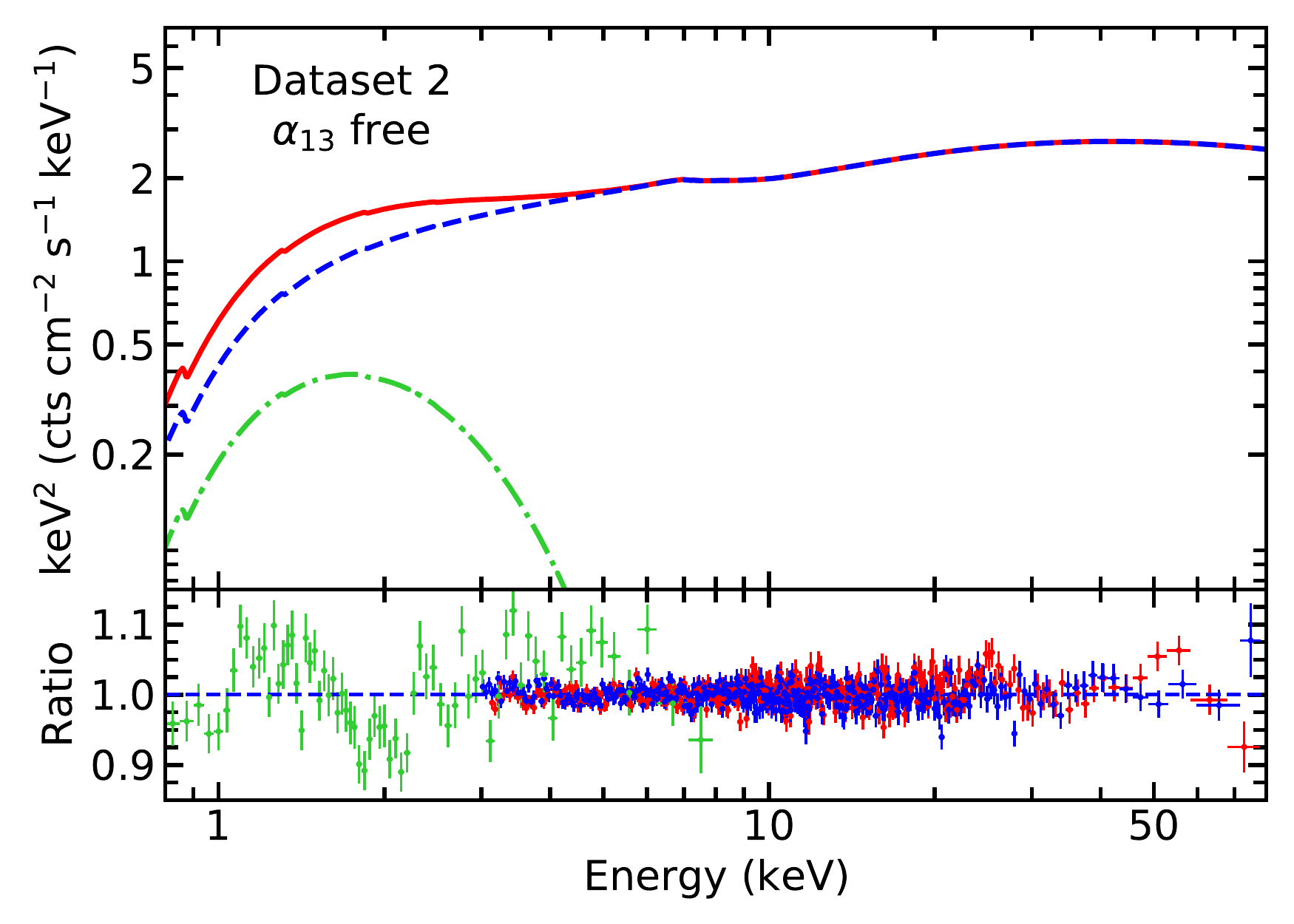} \\ \vspace{0.3cm}
		\includegraphics[scale=0.45]{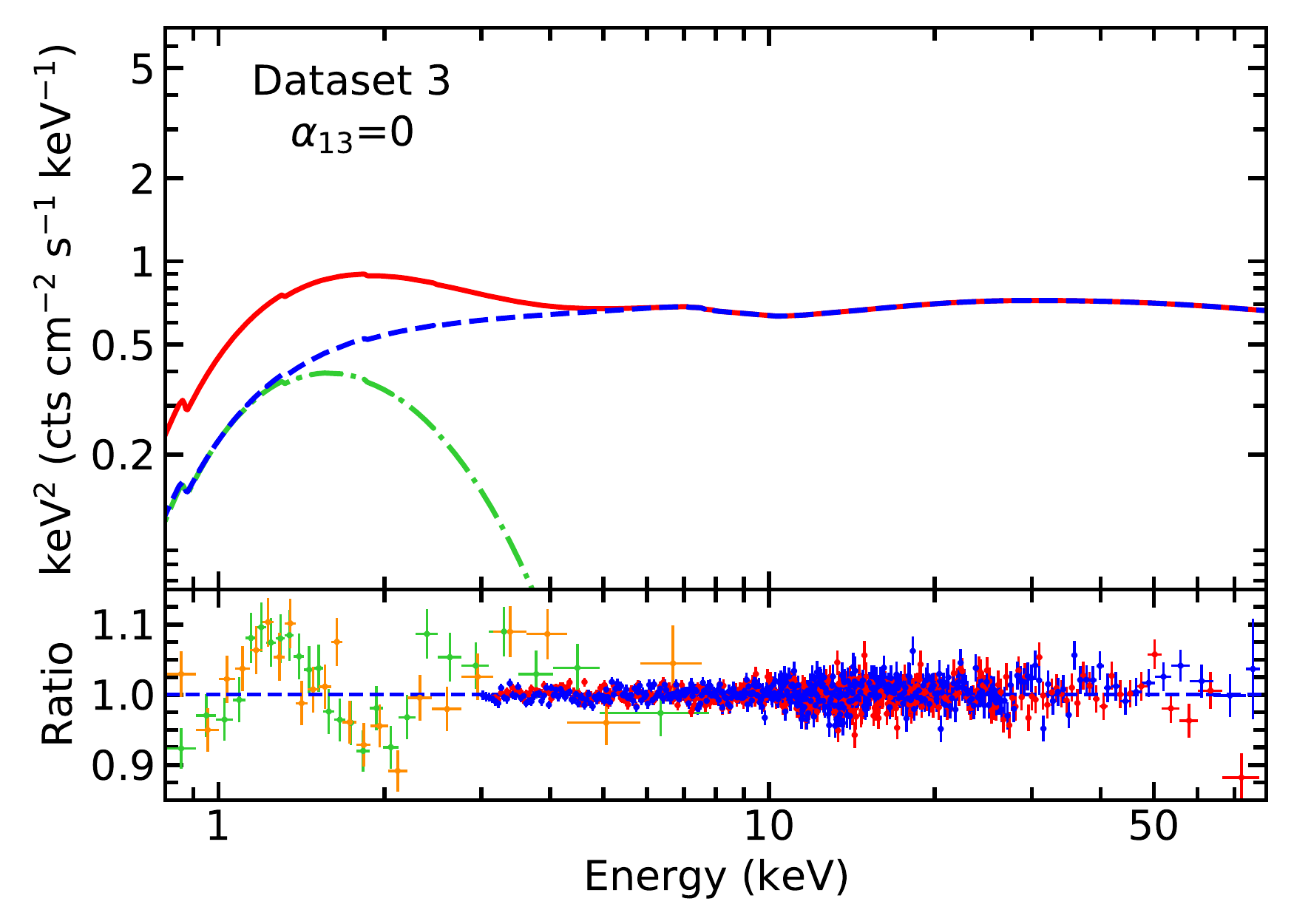}
		\hspace{0.5cm}
		\includegraphics[scale=0.45]{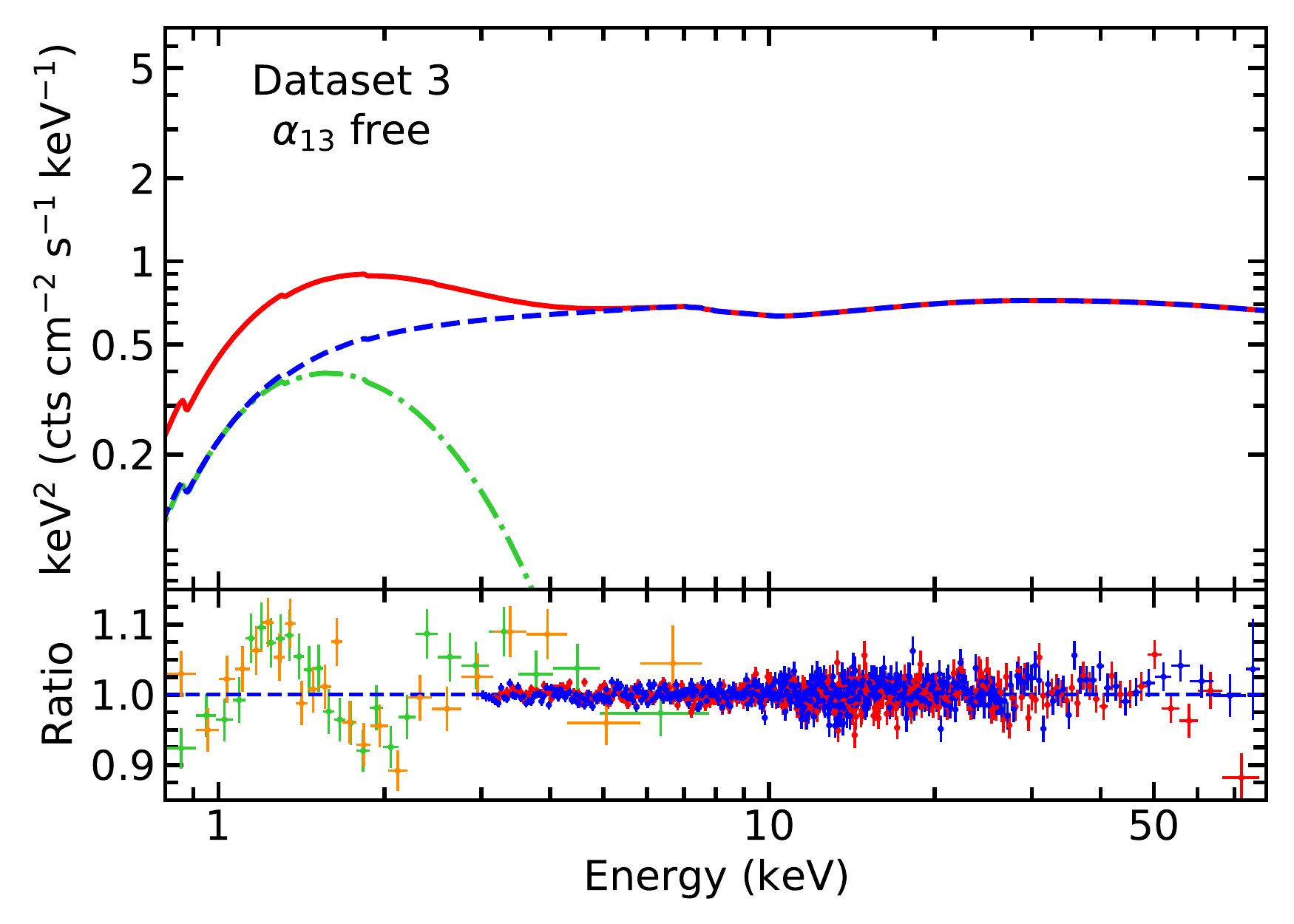}
	\end{center}
	\vspace{-0.2cm}
	\caption{Best-fit models and data to best-fit model ratios when we fit every dataset separately assuming $\alpha_{13} = 0$ (Kerr metric, left panels) and $\alpha_{13}$ free (right panels). In the upper quadrants, the red curves are for the total models, the blue-dashed curves are for the {\tt relxill\_nk} components, and the green-dashed-dotted curves are for the {\tt nkbb} components. In the lower quadrants, red data are for \textsl{NuSTAR}/FPMA, blue data are for \textsl{NuSTAR}/FPMB, green and orange data are for \textsl{Swift}/XRT (in Dataset~2, we only have a \textsl{Swift} snapshot, so there are no orange data). 
	\label{f-bestfit_1}}
\end{figure*}


The analysis of these data was presented in \citet{2019ApJ...887..184T}. They estimated the values of the black hole spin parameter $a_*$ assuming that the spacetime metric around the black hole is described by the Kerr metric. Our study follows the analysis presented in \citet{2019ApJ...887..184T}, but we assume the spacetime metric around the black hole is described by the Johannsen metric \citep{2013PhRvD..88d4002J} with the possible non-vanishing deformation parameter $\alpha_{13}$, while all other deformation parameters are set to zero (the expression of the line element is reported in the appendix of this paper). The models {\tt kerrbb} and {\tt relxill} in \citet{2019ApJ...887..184T} are thus replaced, respectively, by {\tt nkbb} and {\tt relxill\_nk} in our study. Our full XSPEC model is:

\vspace{0.2cm}

\noindent {\tt constant$\times$tbabs$\times$(nkbb+relxill\_nk)}

\vspace{0.2cm}

\noindent {\tt constant} takes the cross-normalization between different instruments into account. {\tt nkbb}~\citep{2019PhRvD..99j4031Z} describes the thermal spectrum of the accretion disk and has the following parameters: the torque at the inner edge of the disk $\eta$, the black hole mass $M$, the black hole distance $D$, the mass accretion rate $\dot{M}$, the inclination angle of the disk with respect to the line of sight of the observer $i$, the index for disk emission $lflag$, the color factor $f_{\rm col}$, the dimensionless black hole spin parameter $a_*$, and the deformation parameter $\alpha_{13}$. In our analysis, we assume limb darkening emission ($lflag=1$) and a vanishing torque at the inner edge of the accretion disk ($\eta=0$). We set the color factor $f_{\rm col}$ to 1.7, which is the most common choice for stellar-mass black holes with an Eddington-scaled disk luminosity of around 10\% and is the same value used in \citet{2019ApJ...887..184T}. The distance of the source is estimated to be $2.4\pm0.4$~kpc~\citep{1994A&A...290..803D} and for the black hole mass there is a lower bound of $4.9$~$M_{\odot}$ \citep{1996A&A...314..123M}.

{\tt relxill\_nk} is an extension of the {\tt relxill} package~\citep{2010MNRAS.409.1534D,2014ApJ...782...76G} to non-Kerr spacetimes. The model has the following parameters: the inner emissivity index $q_{\rm in}$, the outer emissivity index $q_{\rm out}$, and the breaking radius $R_{\rm br}$ to model the emissivity profile of the disk with a broken power-law; the inner and the outer radius of the disk $R_{\rm in}$ and $R_{\rm out}$; the inclination angle of the disk with respect to the line of sight of the observer $i$; the dimensionless black hole spin parameter $a_*$; the deformation parameter $\alpha_{13}$; the photon index $\Gamma$ and the high-energy cutoff $E_{\rm cut}$ to describe the coronal spectrum with a power-law component with a high-energy cutoff; the ionization parameter of the disk $\xi$, the iron abundance $A_{\rm Fe}$ (in units of the Solar iron abundance), and the reflection fraction $R_{\rm ref}$ measuring the relative strength between the reflection component from the disk and the power-law component from the corona. In our analysis, we model the emissivity profile of the disk with a broken power-law in Dataset~1 and Dataset~2, while we assume $q_{\rm in}=q_{\rm out}=3$ for Dataset~3 because these parameters cannot be constrained by the fit. We always assume that the inner edge of the disk is at the innermost stable circular orbit (ISCO) and that the outer edge is at $400~r_{\rm g}$, where $r_{\rm g}$ is the gravitational radius of the black hole. We link the values of the inclination angle $i$, of the black hole spin $a_*$, and of the Johannsen deformation parameter $\alpha_{13}$  between {\tt nkbb} and {\tt relxill\_nk}.

First, we fit the three datasets separately, either assuming the Kerr spacetime with $\alpha_{13}=0$ and with $\alpha_{13}$ free in the fit. In {\tt nkbb}, we freeze the values of the black hole mass to $5$~$M_{\odot}$ and the black hole distance to $2.4$~kpc, ignoring the uncertainty on these two parameters \citep{2019ApJ...887..184T,2021Ap&SS.366...63C}. The results of our fits are summarized in Tab.~\ref{t-bestfit_1} and in Fig.~\ref{f-bestfit_1}. We note that $q_{\rm out}$ is frozen to 1 for Dataset~1 and Dataset~2 because $q_{\rm out}=1$ corresponds to the best-fit value when $q_{\rm out}$ is free but the parameter cannot be constrained.

\begin{figure*}[t]
	\begin{center}
		\includegraphics[scale=0.38]{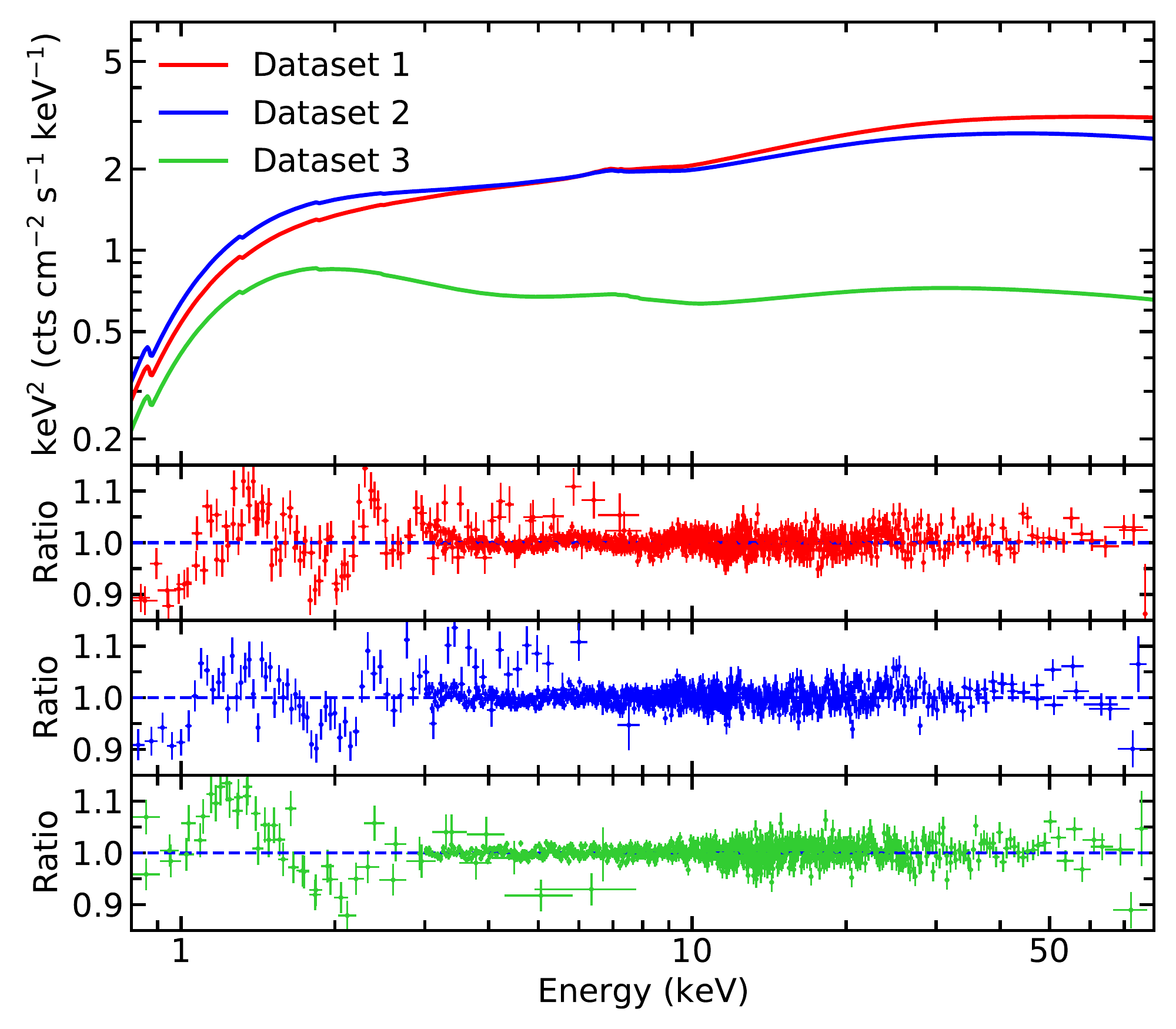}
		\hspace{0.8cm}
		\includegraphics[scale=0.38]{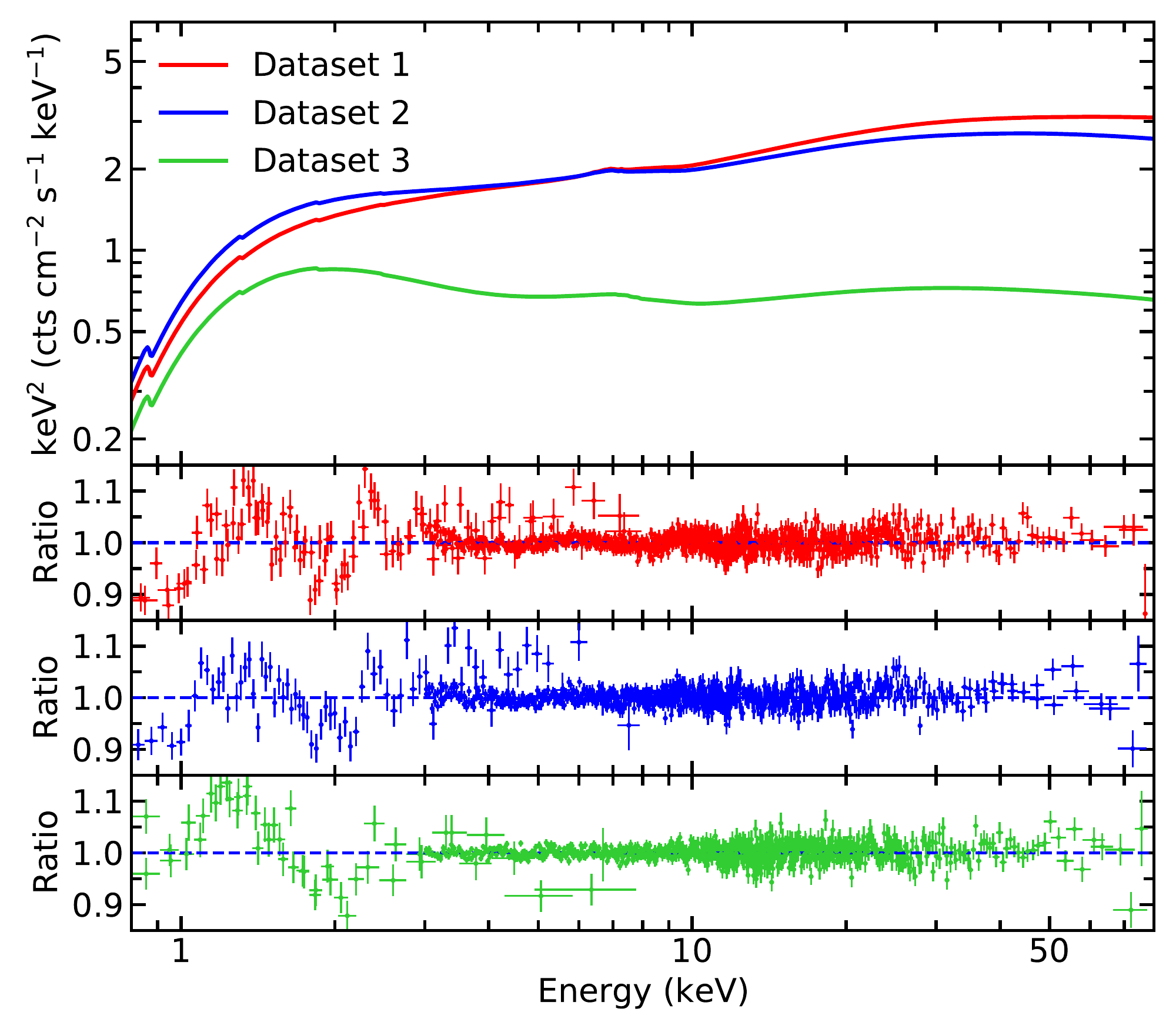}
	\end{center}
	\vspace{-0.2cm}
	\caption{Best-fit models and data to best-fit model ratios when we fit the three datasets together and the values of the black hole mass and distance in {\tt nkbb} are frozen to $M = 5~M_\odot$ and $D = 2.4$~kpc, respectively. In the left panel, we fit the data with $\alpha_{13} = 0$ (Kerr metric). In the right panel, $\alpha_{13}$ is free. 
	\label{ratio_3}}
	\vspace{0.5cm}
\end{figure*}

We then fit all datasets together. We still keep the black hole mass and distance frozen to, respectively, $5$~$M_{\odot}$ and $2.4$~kpc. Since the fits of the individual datasets return a perfectly consistent value for $N_{\rm H}$ in {\tt tbabs}, here we freeze $N_{\rm H}$ to $0.6 \times 10^{22}$~cm$^{-2}$. Even in this case, first we fit the data assuming the Kerr spacetime and then we allow $\alpha_{13}$ to vary in the fit. The results of this joint fit are presented in Tab.~\ref{bestfit_2} and Fig.~\ref{ratio_3}. As for the fits of the single datasets, $q_{\rm out}$ is frozen to 1 for Dataset~1 and Dataset~2 because it cannot be constrained.

Lastly, we include the uncertainties of the black hole mass and distance in our analysis of the three datasets together. We leave the black hole mass in {\tt nkbb} free in the fit but we impose the lower boundary $4.9$~$M_{\odot}$ \citep{1996A&A...314..123M}. The distance $D$ is also free in the fit, but we calculate $\chi^2$ as
\be\label{eq-chi2-d}
\chi^2 + \frac{\left( D - D_* \right)^2}{\sigma_*^2}
\ee
where $D_* = 2.4$~kpc and $\sigma_* = 0.4$~kpc \citep{1994A&A...290..803D}. We fit the data assuming $\alpha_{13} = 0$ and $\alpha_{13}$ free. Table~\ref{bestfit_3} shows the best-fit values.

The results of our spectral analysis will be discussed in the next section.


\begin{table*}[]
	\centering
	{\renewcommand{\arraystretch}{1.3}
	\begin{tabular}{lccc|ccc}
		\hline\hline
		& \multicolumn{3}{c|}{$\alpha_{13}=0$} & \multicolumn{3}{c}{$\alpha_{13}$ free} \\ 
		Parameter & Dataset 1 & Dataset 2 & Dataset 3 & Dataset 1 & Dataset 2 & Dataset 3 \\ \hline 		
		{\tt tbabs}  \\
		$N_{\rm H}$ [$10^{22}$~cm$^{-2}$] & \multicolumn{3}{c|}{$0.6^*$} & \multicolumn{3}{c}{$0.6^*$} \\ \hline 		
		{\tt nkbb} \\ 
		$M$~[$M_{\odot}$]  & \multicolumn{3}{c|}{$5^*$} & \multicolumn{3}{c}{$5^*$}  \\
		$\dot{M}$ [$10^{15}$~g~s$^{-1}$] & $2.64_{-0.25}^{+0.3}$ & $5.18_{-0.26}^{+0.3}$ & $5.37_{-0.18}^{+0.23}$ & $2.51_{-0.18}^{+0.28}$ & $4.96_{-0.3}^{+0.24}$ & $5.19_{-0.24}^{+0.18}$ \\ 
		$D$~[kpc] & \multicolumn{3}{c|}{$2.4^*$} & \multicolumn{3}{c}{$2.4^*$} \\
		$f_{\rm col}$ & \multicolumn{3}{c|}{$1.7^*$} & \multicolumn{3}{c}{$1.7^*$} \\ \hline 
		{\tt relxill\_nk}  \\ 
		$q_{\rm in}$ & $>9.4$ & $7.2_{-0.3}^{+1.6}$ & $3^*$ & $>9.35$ & $7.3_{-1.0}^{+1.1}$ & $3^*$ \\  
		$q_{\rm out}$ & $1^*$ & $1^*$ & $3^*$ & $1^*$ & $1^*$ & $3^*$ \\ 
		$R_{\rm br} [r_{\rm g}]$ & $4.96_{-0.4}^{+0.19}$ & $7.1_{-1.1}^{+0.9}$ & $15^*$ & $5.04_{-0.17}^{+0.3}$ & $7.0_{-1.1}^{+1.8}$ & $15^*$ \\  
		$a_*$ & \multicolumn{3}{c|}{$0.9905\pm0.0013$} & \multicolumn{3}{c}{$>0.989$} \\  
		$i$ [deg] & \multicolumn{3}{c|}{$47.5\pm0.6$} & \multicolumn{3}{c}{$47.8\pm0.8$} \\  
		$ \Gamma $ & $1.675_{-0.004}^{+0.005}$ & $1.708\pm0.007$ & $1.904_{-0.004}^{+0.007}$ & $1.675_{-0.008}^{+0.009}$ & $1.708_{-0.008}^{+0.010}$ & $1.904\pm0.004$ \\
		$\log\xi $ & $3.402_{-0.023}^{+0.029}$ & $3.49_{-0.04}^{+0.05}$ & $3.49_{-0.03}^{+0.04}$ & $3.40\pm0.04$ & $3.48\pm0.05$ & $3.500_{-0.025}^{+0.06}$ \\  
		$ A_{\rm Fe} $ & \multicolumn{3}{c|}{$5.04_{-0.13}^{+0.28}$} & \multicolumn{3}{c}{$5.00_{-0.13}^{+0.4}$} \\ 
		$ E_{\rm cut} $ [keV] & $422_{-49}^{+18}$ & $257_{-20}^{+22}$ & $456_{-46}^{+43}$ & $425_{-22}^{+45}$ & $259_{-20}^{+27}$ & $454_{-69}^{+51}$ \\ 
		$R_{\rm ref}$ & $0.199_{-0.020}^{+0.016}$ & $0.132_{-0.008}^{+0.012}$ & $0.1014_{-0.003}^{+0.0019}$ & $0.201_{-0.017}^{+0.018}$ & $0.134_{-0.010}^{+0.017}$ & $0.1019_{-0.0019}^{+0.006}$ \\ 
		$ \alpha_{13} $ & \multicolumn{3}{c|}{$0^*$} & \multicolumn{3}{c}{$0.083_{-0.10}^{+0.018}$}  \\   
		Norm & $0.0273_{-0.0008}^{+0.0007}$ &  $0.0240\pm0.0006$ & $0.00855_{-0.00021}^{+0.00017}$ & $0.0272_{-0.0008}^{+0.0009}$ & $0.0239_{-0.0006}^{+0.0007}$ & $0.00852_{-0.00021}^{+0.00022}$ \\ \hline  
		Cross-normalization \\
		$C_{\rm FPMB}$ & $1.0135\pm0.0015$ & $1.0147\pm0.0016$ & $1.0159_{-0.0011}^{+0.0012}$ & $1.0135_{-0.0014}^{+0.0015}$ & $1.0147\pm0.0016$ & $1.0159\pm0.0012$ \\ 
		$C_{\rm XRT1}$ & $0.735\pm0.007$ & $0.704\pm0.006$ & $1.023\pm0.013$ & $0.735\pm0.007$ & $0.704\pm0.006$ & $1.021\pm0.011$ \\ 
		$C_{\rm XRT2}$ & $0.955_{-0.010}^{+0.011}$ & & $0.886_{-0.011}^{+0.012}$ & $0.955\pm0.010$ & & $0.884\pm0.010$ \\ 
		$\chi^2/\nu $ & \multicolumn{3}{c|}{$10735.06/9691$} & \multicolumn{3}{c}{$10733.39/9690$} \\ 
		& \multicolumn{3}{c|}{$=1.107735$} & \multicolumn{3}{c}{$=1.107677$} \\ 
		\hline\hline
	\end{tabular}
	}
	\caption{\rm Best-fit values when we fit the three datasets together and we freeze the values of the black hole mass and distance in {\tt nkbb}. The reported uncertainties correspond to the 90\% confidence level for one relevant parameter ($\Delta\chi^2 = 2.71$). $^*$ indicates that the parameter is frozen in the fit. $\xi$ is in units of erg~cm~s$^{-1}$. See the text for more details.
	\label{bestfit_2}}
\end{table*}


\begin{table*}[]
	\centering
	{\renewcommand{\arraystretch}{1.3}
	\begin{tabular}{lccc|ccc}
		\hline\hline
		Model & \multicolumn{3}{c|}{$\alpha_{13}=0$} & \multicolumn{3}{c}{$\alpha_{13}$ free} \\ 
		& Dataset 1 & Dataset 2 & Dataset 3 & Dataset 1 & Dataset 2 & Dataset 3 \\ \hline 		
		{\tt tbabs}  \\
		$N_{\rm H}$ [$10^{22}$~cm$^{-2}$] & \multicolumn{3}{c|}{$0.6^*$} & \multicolumn{3}{c}{$0.6^*$} \\ \hline 		
		{\tt nkbb} \\ 
		$M$~[$M_{\odot}$]  & \multicolumn{3}{c|}{$5.12_{\rm -(B)}^{+0.18}$} & \multicolumn{3}{c}{$4.97_{\rm -(B)}^{+1.41}$}  \\
		$\dot{M}$ [$10^{15}$~g~s$^{-1}$] & $2.58_{-0.29}^{+0.42}$ & $5.13_{-0.5}^{+0.12}$ & $5.33_{-0.17}^{+1.6}$ & $2.56_{-0.5}^{+1.1}$ & $4.99_{-0.8}^{+2.6}$ & $5.2_{-0.6}^{+3.0}$ \\ 
		$D$~[kpc] & \multicolumn{3}{c|}{$2.39_{-0.13}^{+0.16}$} & \multicolumn{3}{c}{$2.41_{-0.17}^{+0.6}$} \\
		$f_{\rm col}$ & \multicolumn{3}{c|}{$1.7^*$} & \multicolumn{3}{c}{$1.7^*$} \\ \hline 
		{\tt relxill\_nk}  \\ 
		$q_{\rm in}$ & $>9.62$ & $7.22_{-0.17}^{+0.7}$ & $3^*$ & $>9.27$ & $7.4_{-0.9}^{+1.1}$ & $3^*$ \\  
		$q_{\rm out}$ & $1^*$ & $1^*$ & $3^*$ & $1^*$ & $1^*$ & $3^*$ \\ 
		$R_{\rm br} [r_{\rm g}]$ & $4.96_{-0.19}^{+0.06}$ & $7.01_{-0.6}^{+0.24}$ & $15^*$ & $5.03_{-0.17}^{+0.3}$ & $7.0_{-1.3}^{+1.9}$ & $15^*$ \\  
		$a_*$ & \multicolumn{3}{c|}{$0.9912_{-0.0021}^{+0.0011}$} & \multicolumn{3}{c}{$>0.989$} \\  
		$i$ [deg] & \multicolumn{3}{c|}{$47.9_{-0.8}^{+0.5}$} & \multicolumn{3}{c}{$47.8\pm0.9$} \\  
		$ \Gamma $ & $1.6760_{-0.003}^{+0.0016}$ & $1.708\pm0.003$ & $1.904_{-0.003}^{+0.0029}$ & $1.675_{-0.008}^{+0.010}$ & $1.708_{-0.008}^{+0.010}$ & $1.904\pm0.008$ \\
		$\log\xi $ & $3.398_{-0.012}^{+0.021}$ & $3.484_{-0.020}^{+0.05}$ & $3.498_{-0.012}^{+0.019}$ & $3.401_{-0.03}^{+0.026}$ & $3.48_{-0.05}^{+0.04}$ & $3.50_{-0.04}^{+0.06}$ \\  
		$ A_{\rm Fe} $ & \multicolumn{3}{c|}{$5.03_{-0.13}^{+0.24}$} & \multicolumn{3}{c}{$5.00_{-0.14}^{+0.5}$} \\ 
		$ E_{\rm cut} $ [keV] & $431_{-26}^{+14}$ & $258_{-12}^{+5}$ & $459_{-27}^{+51}$ & $419_{-40}^{+78}$ & $258_{-23}^{+47}$ & $455_{-48}^{+67}$ \\ 
		$R_{\rm ref}$ & $0.203_{-0.023}^{+0.009}$ & $0.133_{-0.006}^{+0.007}$ & $0.1019_{-0.0012}^{+0.0015}$ & $0.198_{-0.016}^{+0.026}$ & $0.133_{-0.013}^{+0.029}$ & $0.1016_{-0.0020}^{+0.006}$ \\ 
		$ \alpha_{13} $ & \multicolumn{3}{c|}{$0^*$} & \multicolumn{3}{c}{$0.090_{-0.12}^{+0.012}$}  \\   
		Norm & $0.0273_{-0.0007}^{+0.0004}$ &  $0.0240\pm0.0003$ & $0.00853_{-0.00012}^{+0.00013}$ & $0.0272_{-0.0008}^{+0.0009}$ & $0.0239\pm0.0007$ & $0.00853\pm0.00022$ \\ \hline  
		Cross-normalization \\
		$C_{\rm FPMB}$ & $1.0135\pm0.0015$ & $1.0147\pm0.0016$ & $1.0159_{-0.0011}^{+0.0012}$ & $1.0135\pm0.0015$ & $1.0147\pm0.0016$ & $1.0159_{-0.0011}^{+0.0012}$ \\ 
		$C_{\rm XRT1}$ & $0.735_{-0.006}^{+0.007}$ & $0.703\pm0.006$ & $1.022\pm0.013$ & $0.703\pm0.007$ & $0.704\pm0.006$ & $1.022_{-0.015}^{+0.014}$ \\ 
		$C_{\rm XRT2}$ & $0.955\pm0.010$ & & $0.885\pm0.012$ & $0.955_{-0.010}^{+0.011}$ & & $0.885\pm0.013$ \\ 
		$\chi^2/\nu $ & \multicolumn{3}{c|}{$10735.03/9690$} & \multicolumn{3}{c}{$10733.36/9689$} \\ 
		& \multicolumn{3}{c|}{$=1.107846$} & \multicolumn{3}{c}{$=1.107787$} \\ 
		\hline\hline
	\end{tabular}
	}
	\caption{\rm Best-fit values when we fit the three datasets together and we leave free the values of the black hole mass and distance in {\tt nkbb}. The reported uncertainties correspond to the 90\% confidence level for one relevant parameter ($\Delta\chi^2 = 2.71$). $^*$ indicates that the parameter is frozen in the fit. (B) in the uncertainty of $M$ indicates that we reach the lower boundary of the parameter range (4.9~$M_\odot$) before the 90\% confidence level limit. $\xi$ is in units of erg~cm~s$^{-1}$. See the text for more details.
	\label{bestfit_3}}
\end{table*}


\section{Discussion and conclusions}\label{s-dis}

In the previous section, we have analyzed \textsl{NuSTAR} and \textsl{Swift} spectra of the Galactic black hole GRS~1716--249 in the hard-intermediate state during the outburst of 2016-2017. The data were analyzed in \citet{2019ApJ...887..184T} assuming that the spacetime metric around the black hole is described by the Kerr solution. In our study, we have analyzed the same observations employing the latest versions of {\tt nkbb} and {\tt relxill\_nk}, which are the state-of-the-art to test the Kerr nature of an accreting black hole from the analysis of, respectively, the thermal spectrum and the reflection features of the black hole accretion disk.

Our results for $\alpha_{13} = 0$ (Kerr metric) are consistent with those reported in \citet{2019ApJ...887..184T}, where the authors report only the results of the individual fits of the three datasets. The value of the iron abundance $A_{\rm Fe}$ is high, but this is a well-known issue in the study of the relativistic reflection features of stellar-mass and supermassive black holes and the exact origin is currently unknown; see, e.g., \citet{2021SSRv..217...65B} and reference therein for possible explanations.

\begin{figure*}[t]
	\begin{center}
		\includegraphics[scale=0.33,trim={2cm 1cm 1cm 2cm},clip]{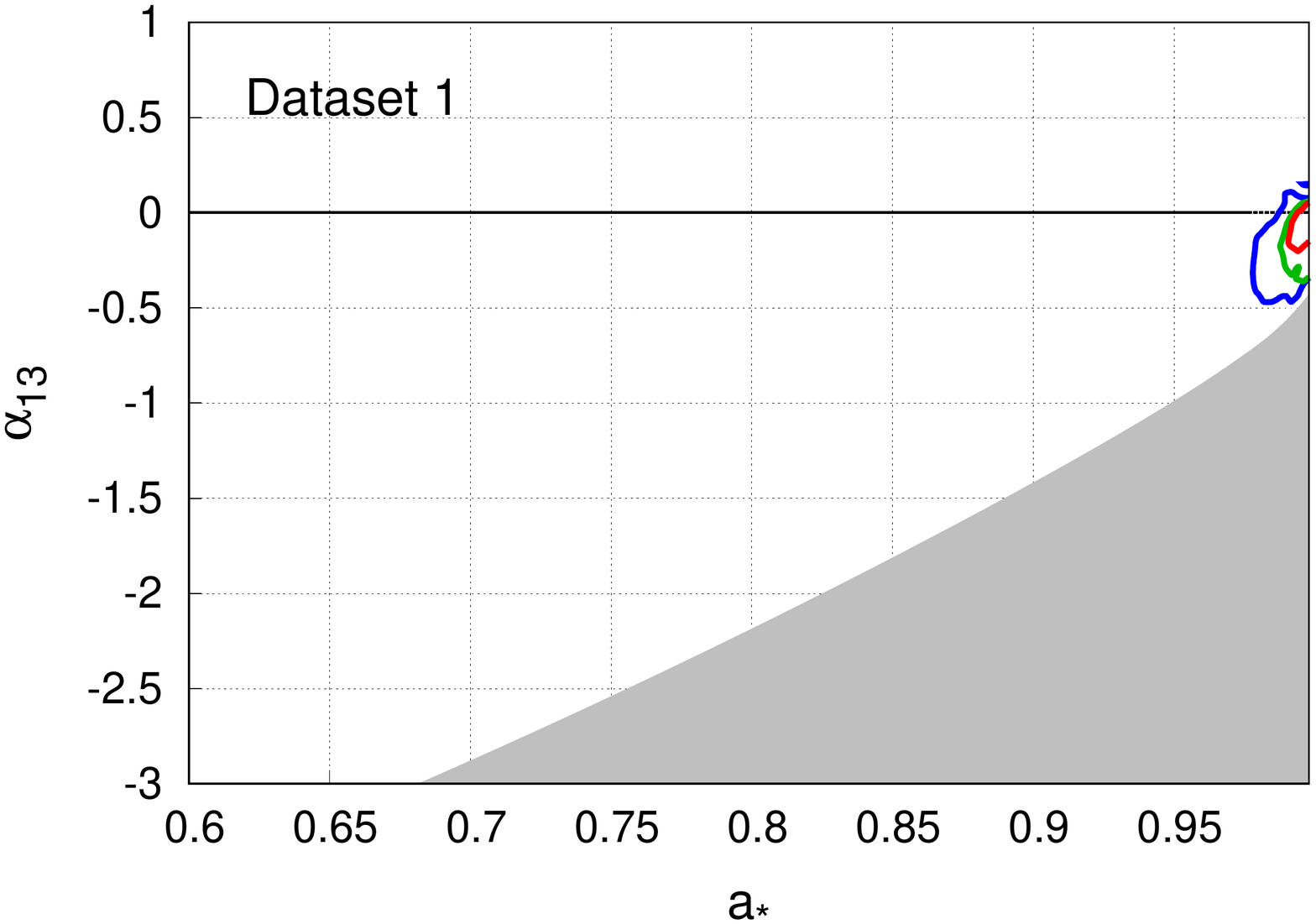}
		\hspace{0.5cm}
		\includegraphics[scale=0.33,trim={2cm 1cm 1cm 2cm},clip]{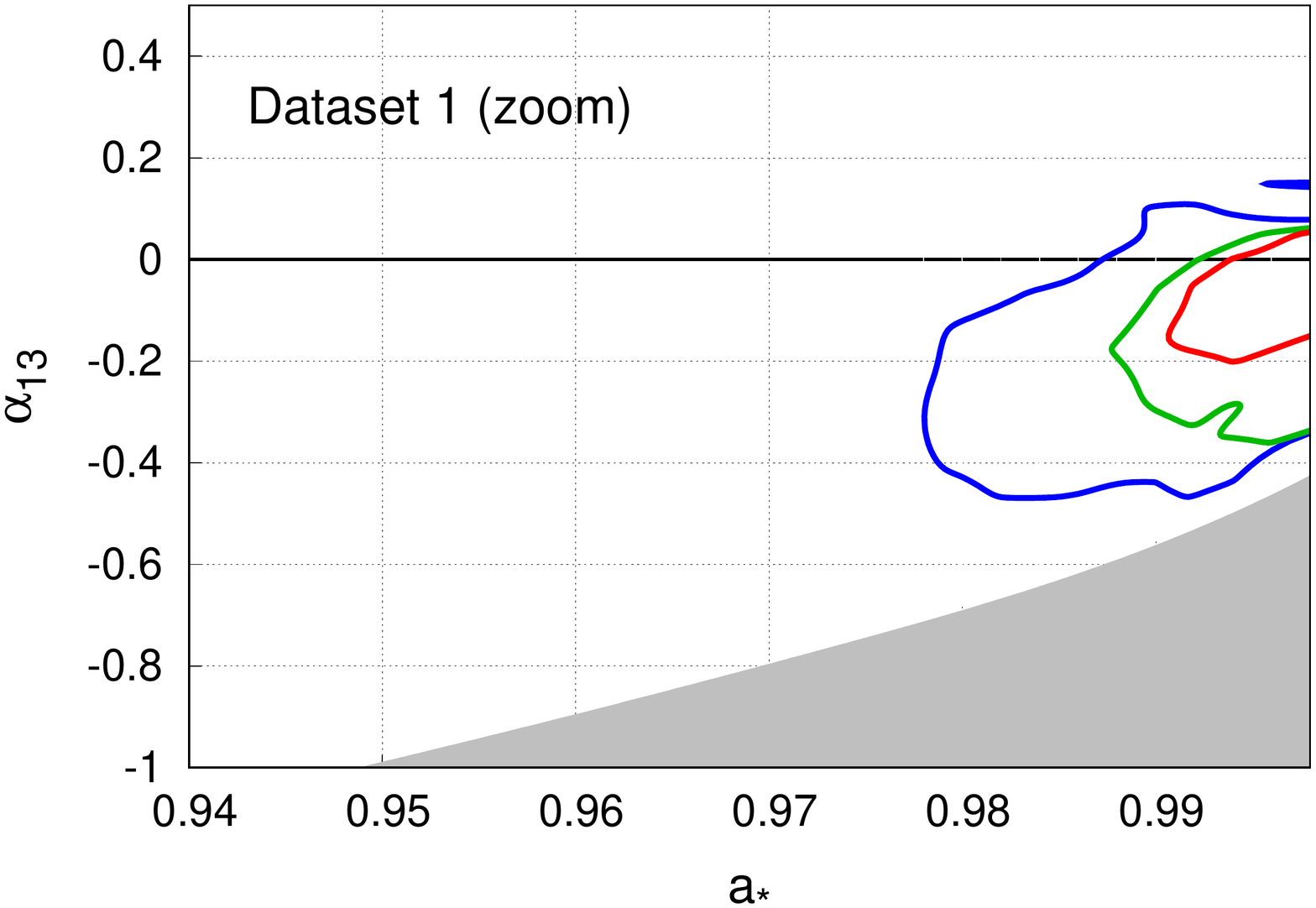} \\
		\includegraphics[scale=0.33,trim={2cm 1cm 1cm 2cm},clip]{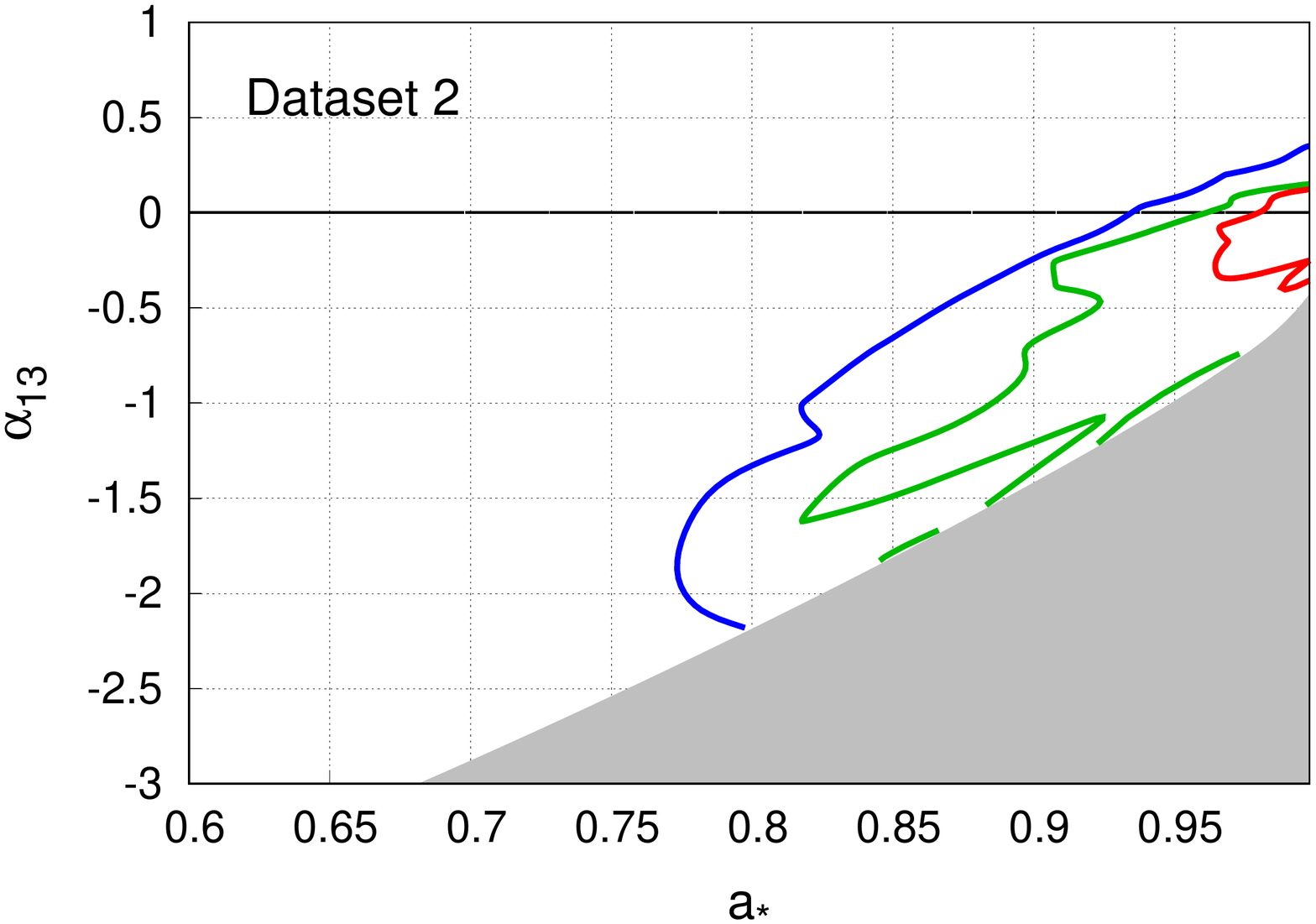}
		\hspace{0.5cm}
		\includegraphics[scale=0.33,trim={2cm 1cm 1cm 2cm},clip]{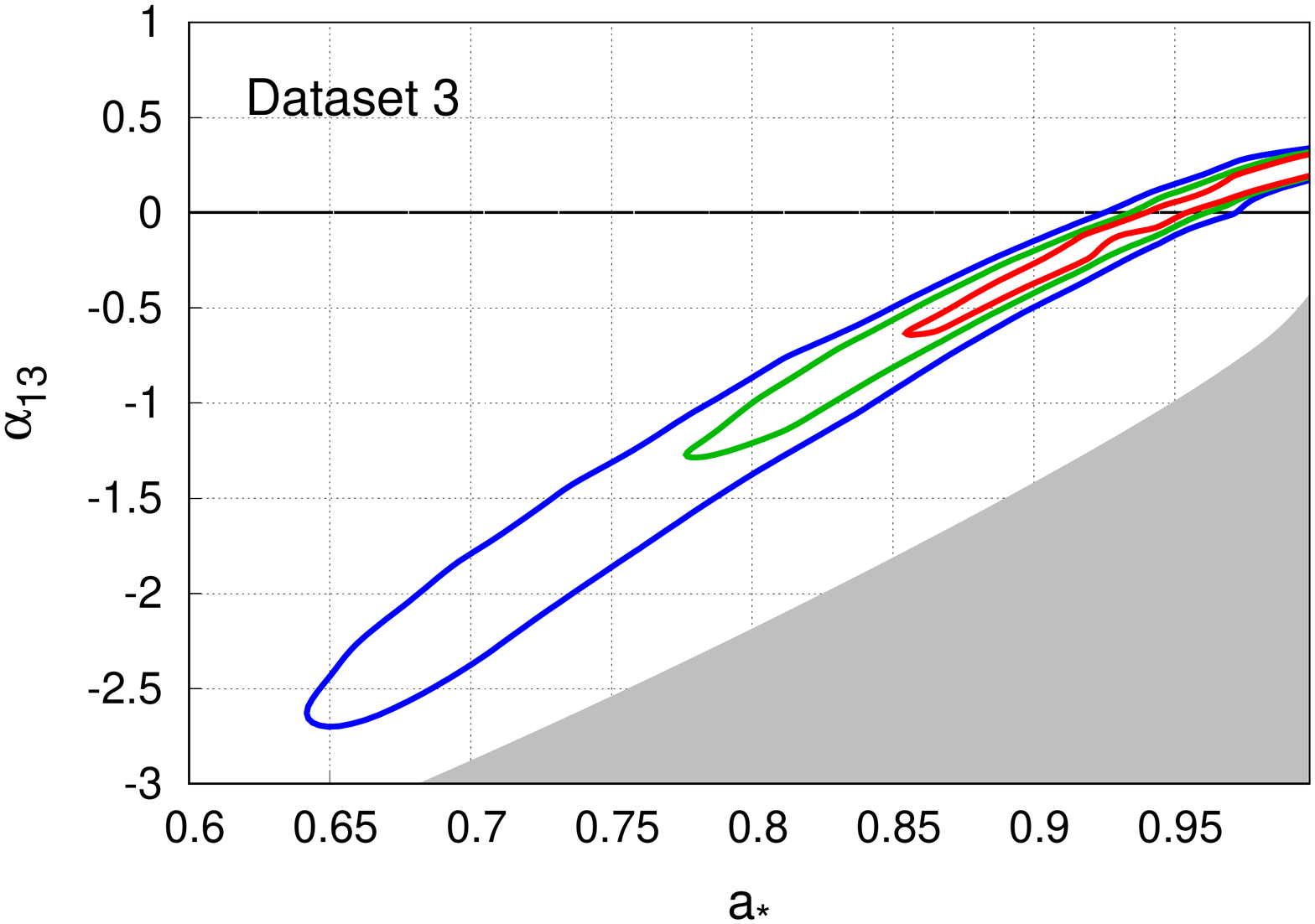} 
	\end{center}
	\vspace{-0.7cm}
	\caption{Constraints on the spin parameter $a_*$ and the Johannsen deformation parameter $\alpha_{13}$ when we fit every dataset separately. The top-right panel is an enlargement of the top-left panel for Dataset~1. The red, green, and blue curves represent, respectively, the 68\%, 90\%, and 99\% confidence level limits for two relevant parameters ($\Delta\chi^2 = 2.30$, 4.61, and 9.21, respectively). The thicker horizontal line at $\alpha_{13}=0$ marks the Kerr solution. The gray region is ignored in our analysis because the spacetime has pathological properties there; see Eq.~(\ref{eq-app-c}) in the appendix.
       \label{contour_1}}
       \vspace{0.5cm}
       	\begin{center} 
	\includegraphics[scale=0.33,trim={2cm 1cm 1cm 2cm},clip]{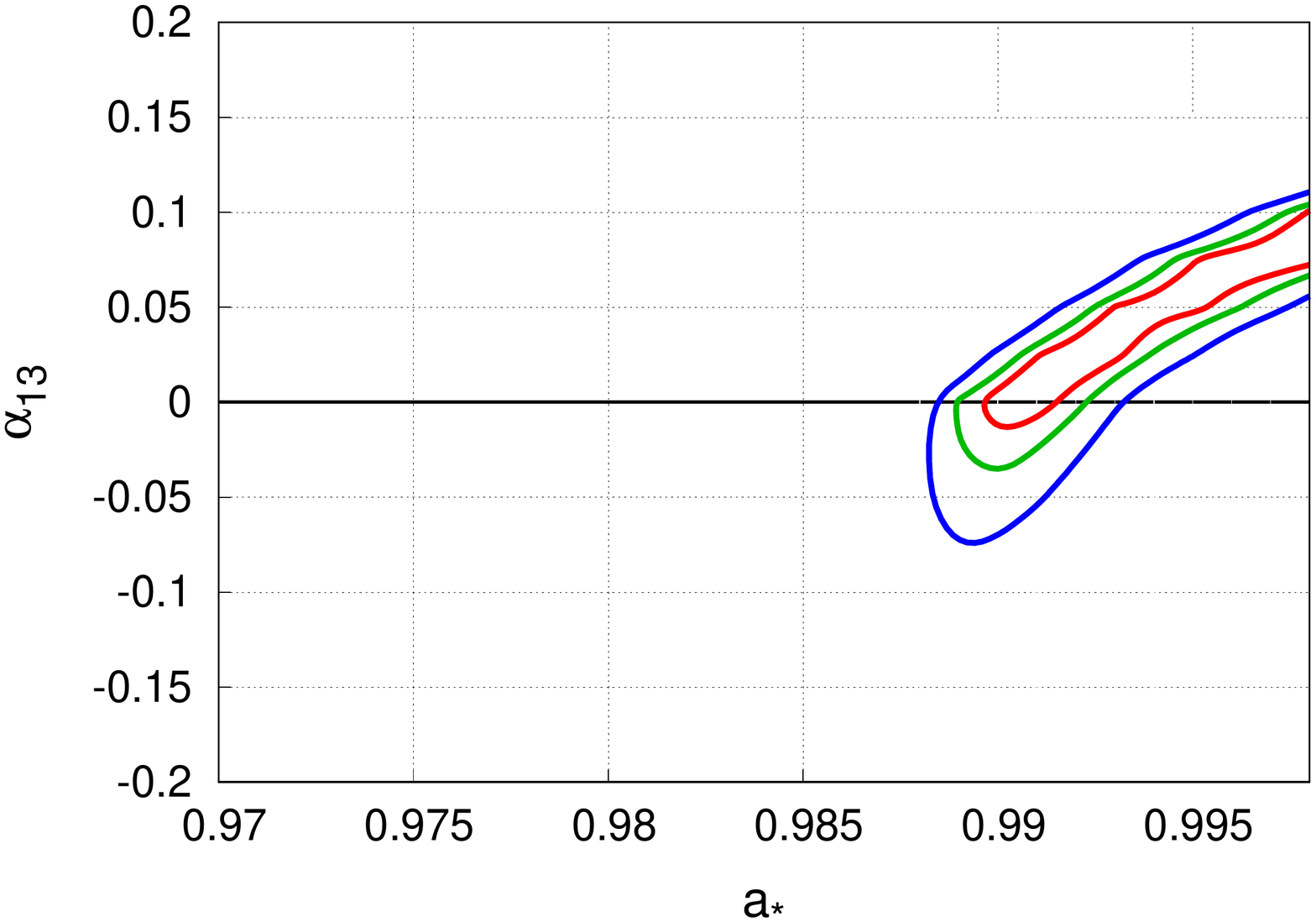}
	\hspace{0.5cm}
	\includegraphics[scale=0.33,trim={2cm 1cm 1cm 2cm},clip]{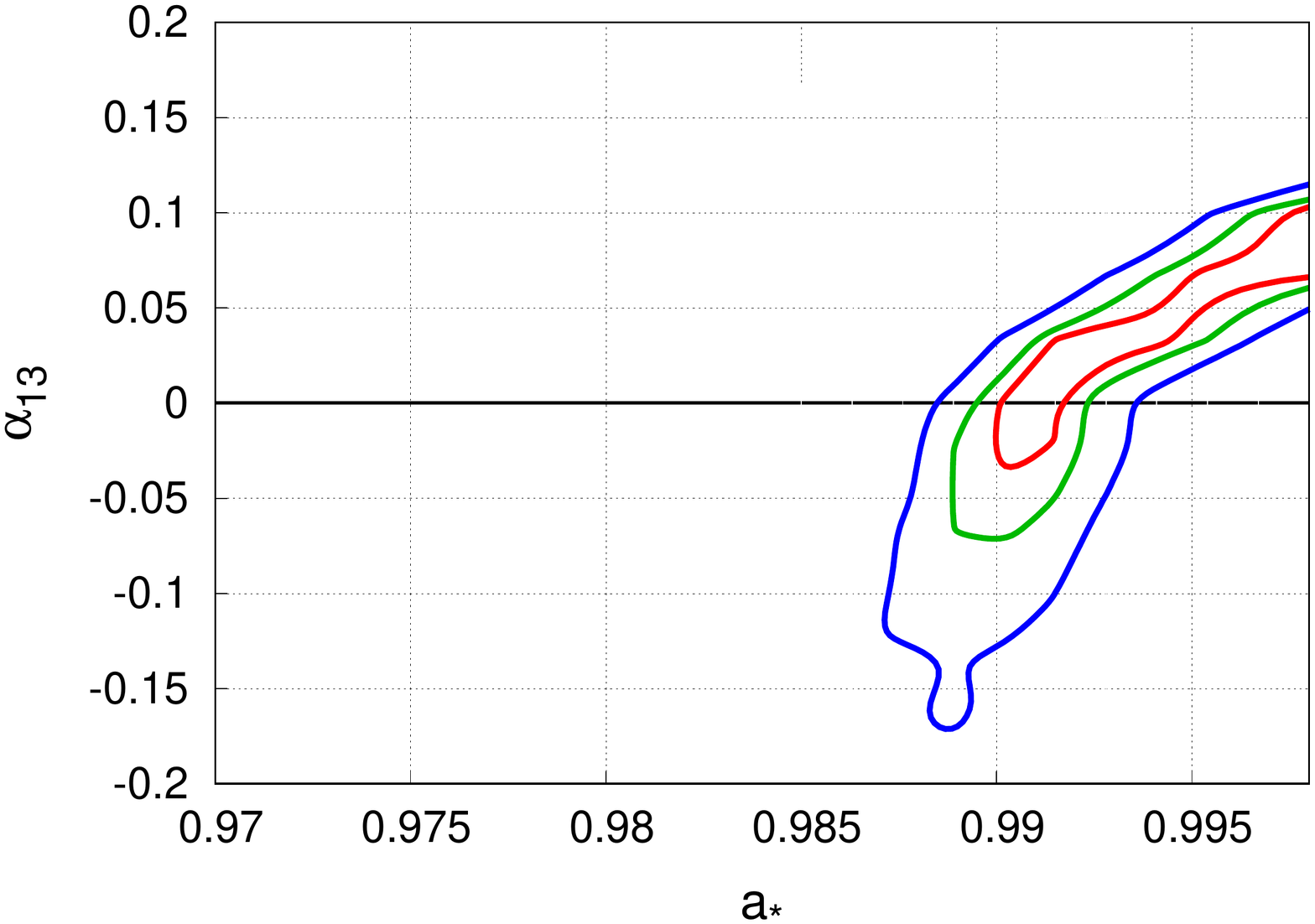}
	\end{center}
	\vspace{-0.7cm}
	\caption{Constraints on the spin parameter $a_*$ and the Johannsen deformation parameter $\alpha_{13}$ when we fit the three datasets together. In the left panel, the black hole mass $M$ and the distance $D$ are frozen in {\tt nkbb}. In the right panel, these two parameters are free. The red, green, and blue curves represent, respectively, the 68\%, 90\%, and 99\% confidence level limits for two relevant parameters ($\Delta\chi^2 = 2.30$, 4.61, and 9.21, respectively). The thicker horizontal line at $\alpha_{13}=0$ marks the Kerr solution.
       \label{contour_disun_massfree_joint}}
\end{figure*}

The main result of this paper is the constraint on the deformation parameter $\alpha_{13}$, which is obtained by fitting simultaneously the thermal spectrum and the reflection features of the source. When we fit the three datasets separately, we obtain the constraints on the black hole spin $a_*$ and on the Johannsen deformation parameter $\alpha_{13}$ shown in Fig.~\ref{contour_1}. The constraints from Dataset~1 are much stronger than those obtained from the other two datasets. For Dataset~3, the plot in Fig.~\ref{contour_1} shows the typical banana shape of the confidence level curves, which appears when there is a strong degeneracy between the two parameters. In Dataset~1, the constraints are mainly determined by the analysis of the reflection features. In Dataset~3, we have instead that the constraints are mainly inferred from the thermal component of the disk, where it is well-known that the spectrum is degenerate between $a_*$ and $\alpha_{13}$ and the shape of the confidence level curves in Fig.~\ref{contour_1} is similar to that found in \citet{2020ApJ...897...84T}, where we analyzed with {\tt nkbb} only thermally-dominated spectra of the black hole in LMC~X-1.


\begin{figure*}[t]
	\begin{center}
	\includegraphics[scale=0.37,trim={0cm 0cm 0cm 0cm},clip]{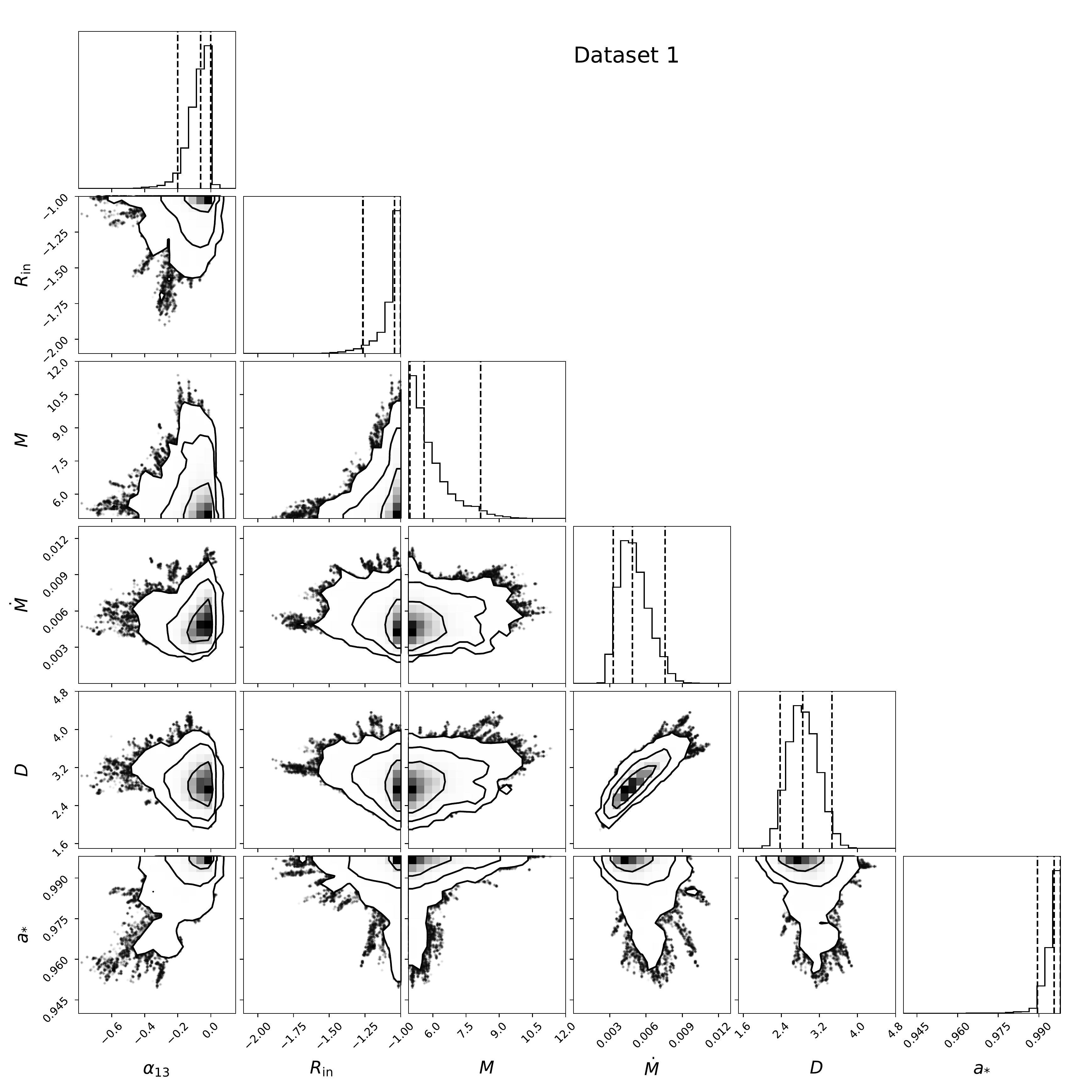}
	\end{center}
	\vspace{-0.3cm}
	\caption{{Correlations among the deformation parameter $\alpha_{13}$, the inner edge of the accretion disk $R_{\rm in}$ (in units of ISCO radius and with a minus sign, which is the output from {\tt relxill\_nk}), the black hole mass $M$ (in Solar mass units), the mass accretion rate $\dot{M}$ (in units $10^{15}$~g~s$^{-1}$), the black hole distance $D$ (in kpc), and the black hole spin parameter $a_*$ for Dataset~1. The dashed vertical lines denote the 5th, 50th, and 95th percentiles for individual parameters.}
       \label{mcmc1}}
\end{figure*}

\begin{figure*}[t]
	\begin{center}
	\includegraphics[scale=0.37,trim={0cm 0cm 0cm 0cm},clip]{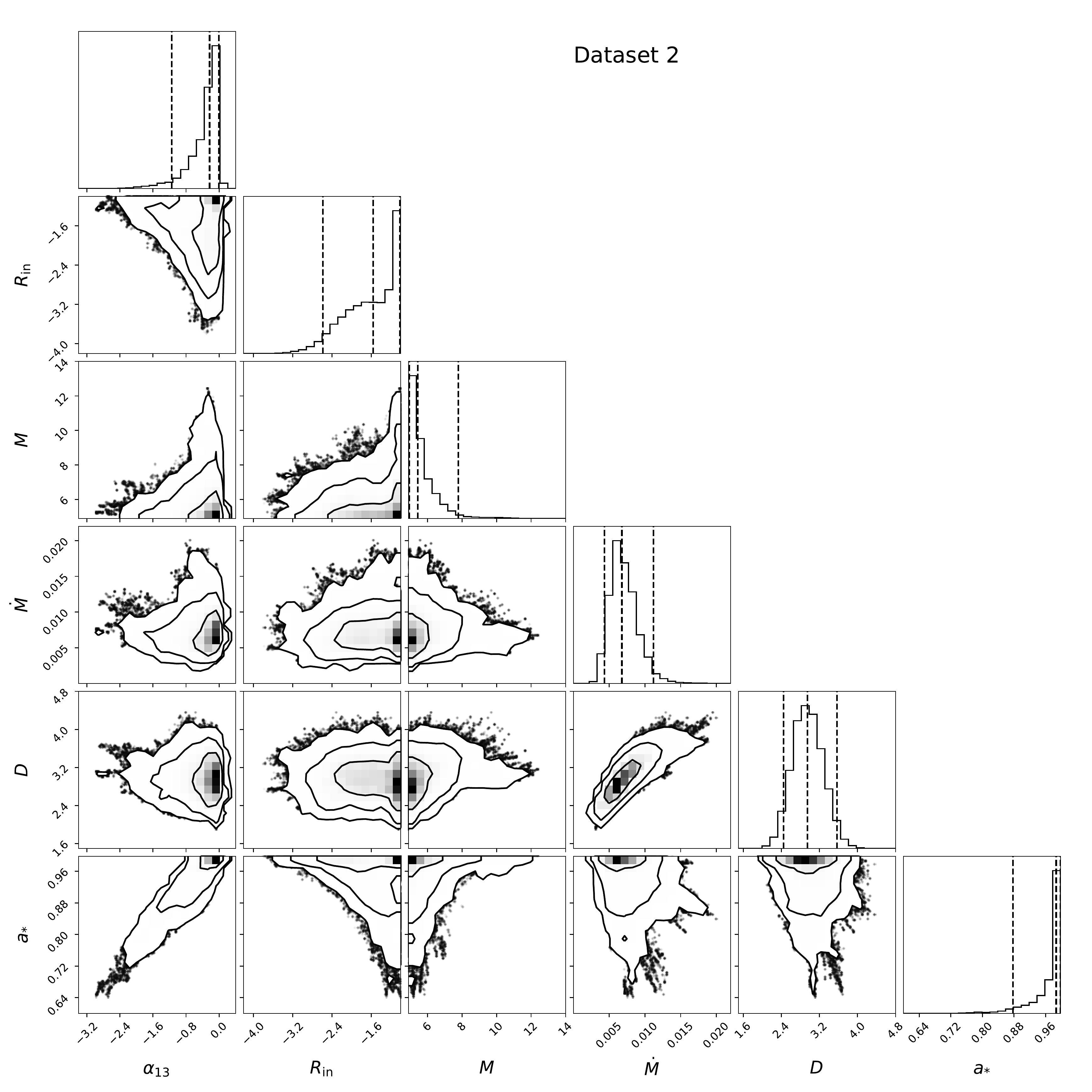}
	\end{center}
	\vspace{-0.3cm}
	\caption{{As in Fig.~\ref{mcmc1} for Dataset~2}
       \label{mcmc2}}
\end{figure*}

\begin{figure*}[t]
	\begin{center}
	\includegraphics[scale=0.37,trim={0cm 0cm 0cm 0cm},clip]{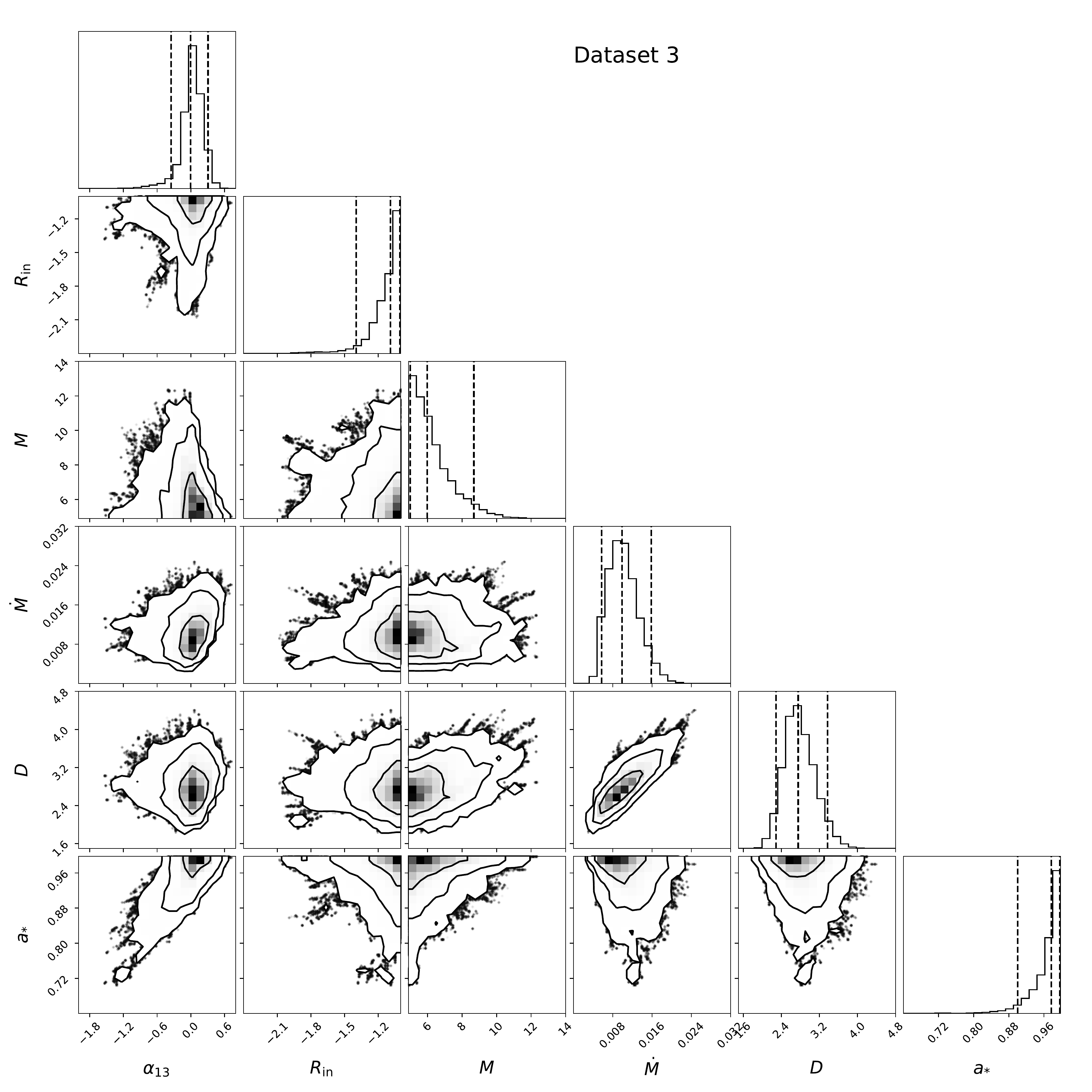}
	\end{center}
	\vspace{-0.3cm}
	\caption{{As in Fig.~\ref{mcmc1} for Dataset~3}
       \label{mcmc3}}
\end{figure*}

When we fit the three datasets together, our constraints on the deformation parameter $\alpha_{13}$ get stronger. For $M$ and $D$ frozen in {\tt nkbb}, we obtain the constraints on $a_*$ and $\alpha_{13}$ shown in the left panel of Fig.~\ref{contour_disun_massfree_joint}. When we leave $M$ and $D$ free in {\tt nkbb}, we impose $M \ge 4.9~M_\odot$, and we include the uncertainty on the measurement of the distance by using the $\chi^2$ in Eq.~(\ref{eq-chi2-d}), we find the constraints in the right panel in Fig.~\ref{contour_disun_massfree_joint}, which are only slightly worse than those in the left panel with $M$ and $D$ fixed in the fit. In the end, our measurement of $\alpha_{13}$ from the fit of the three datasets together and including the uncertainties on the black hole mass and distance is
\be
\alpha_{13} = 0.090_{-0.12}^{+0.012} \quad {\rm (90\% C.L.)} \, .
\ee

We note that in all our fits we assumed the inner edge of the accretion disk, $R_{\rm in}$, to be at the ISCO radius, as suggested in \citet{2019MNRAS.482.1587B}, and thus $R_{\rm in}$ is determined by the values of $a_*$ and $\alpha_{13}$. However, the analysis of the reflection spectra of GRS~1716--249 by \citet{2020MNRAS.492.1947J} shows that the disk may be slightly truncated. {We thus repeated the analysis of the three datasets separately assuming $R_{\rm in}$ free when the deformation parameter $\alpha_{13}$, the black hole mass $M$, and the black hole distance $D$ are also free in the fit. For every dataset, we performed a Markov Chain Monte-Carlo (MCMC) analysis with 100 walkers, 4.8 million steps, and a burn-in of 0.8 million steps. The corner plots showing the correlations among $\alpha_{13}$, $R_{\rm in}$, $M$, $\dot{M}$, $D$, and $a_*$ are in Fig.~\ref{mcmc1}, Fig.~\ref{mcmc2}, and Fig.~\ref{mcmc3} for, respectively, Dataset~1, Dataset~2, and Dataset~3. $R_{\rm in}$ is measured in units of $R_{\rm ISCO}$ (with a minus sign, which is the output of {\tt relxill\_nk}) and is always consistent with the ISCO radius. Its measurement is not correlated to the other model parameters, but this is just because these \textsl{NuSTAR} spectra require that $R_{\rm in}$ is as small as possible, so it should be at the ISCO (our model does not allow that the inner edge is inside the ISCO). From these corner plots, we also note that the black hole distance $D$ and the mass accretion rate $\dot{M}$ are always correlated, which is perfectly understandable because these two parameters set the normalization of the thermal component. In Dataset~2 and Dataset~3, we see a clear correlation even between the estimate of the black hole spin $a_*$ and the deformation parameter $\alpha_{13}$, while this is not the case for Dataset~1 where the reflection features are stronger and it is possible to break this parameter degeneracy.}

\subsection{Method justification}

{In the past two decades, there has been a substantial amount of work to understand the potential biases and shortcomings of the continuum-fitting and the iron-line methods employed in our analysis, even if most of these studies have been devoted to investigate the impact on spin measurements assuming the Kerr background \citep[for a review, see][]{2014SSRv..183..295M,2021SSRv..217...65B}.}

{For the continuum-fitting method, a crucial point is the validity of the Novikov-Thorne model for the description of the accretion disk \citep{1973blho.conf..343N,1974ApJ...191..499P}. The time-averaged structure of the disk is governed by the conservation of rest-mass, energy, and angular momentum. However, the model requires one more assumption that does not follow directly from a conservation law but enters as a boundary condition: the viscous torque vanishes inside the ISCO radius. With such an assumption, there is no emission inside the ISCO and the peak of the disk luminosity is at a radius a bit larger than the ISCO. However, magnetic stresses can invalidate this assumption and cause strong torques at the ISCO and in the plunging region~\citep{1999ApJ...522L..57G,1999ApJ...515L..73K}. GRMHD simulations of geometrically thin disks in the Kerr background confirm the presence of non-vanishing magnetic stresses inside the ISCO \citep{2010MNRAS.408..752P,2011MNRAS.414.1183K,2012MNRAS.424.2504Z}. The emission does not vanish inside the ISCO and the peak of the disk luminosity is at a radius smaller than that predicted in the Novikov-Thorne model: both effects lead to overestimate the black hole spin parameter. However, the conclusion of those studies is that these modeling uncertainties are subdominant with respect to the current errors in $a_*$ induced by the uncertainties on the black hole mass and distance\footnote{We note that \citet{2010ApJ...711..959N} employ a different magnetic field configuration and find larger deviations from the predictions of the Novikov-Thorne model.}. In the future, the Novikov-Thorne model may be replaced by simulation-based models \citep{2012MNRAS.420..684P} in order to reduce these modeling biases.}

{While the emissivity profile of the thermal spectrum is determined by the accretion disk model, in the case of the reflection spectrum it is determined by the coronal geometry, which is unknown. Emissivity profiles of coronae of arbitrary geometry are commonly modeled with a power-law, a broken power-law, or even a twice broken power-law. The emissivity profiles of specific coronal geometries have been calculated by several authors \citep[see, e.g.,][]{2003MNRAS.344L..22M,2012MNRAS.424.1284W,2013MNRAS.430.1694D,2017MNRAS.472.1932G,2020arXiv201207469R}. It turns out that a broken power-law or twice broken power-law emissivity profile describes quite accurately most coronal geometries, with the result that even an accurate measurement of the sole emissivity profile is not enough to to infer the coronal geometry \citep{2017MNRAS.472.1932G}. For example, \citet{2021ApJ...913...79T} analyze high-quality \textsl{NuSTAR} spectra of Galactic black holes with {\tt relxill\_nk} and show that, whenever a lamppost emissivity profile provides the best fit, the fit with a phenomenological broken power-law is only marginally worse and the estimate of the model parameters is always consistent with the fit with the lamppost emissivity profile, while the contrary is not true (that is, if the broken power-law model provides the best fit, the fit with the lamppost model may be significantly worse and provides inconsistent estimate of the model parameters). The interpretation of \citet{2021ApJ...913...79T} is that the phenomenological broken power-law model is flexible enough that it can describe well the emissivity profile generated by a lamppost corona for the quality of current data. \citet{2017MNRAS.472.1932G} arrive at a similar conclusion from the analysis of a number of different coronal geometries, whose emissivity profiles can be approximated well by a twice broken power-law.}

\subsection{Current constraints on $\alpha_{13}$ in the literature}

The constraint on $\alpha_{13}$ obtained for GRS~1716--249 in the present work can be compared with the constraints on $\alpha_{13}$ for stellar-mass black holes reported in the literature. Such a comparison is shown in Tab.~\ref{t-ssummary}, where we see that constraints have been obtained from 16~sources by using the continuum-fitting method, the iron line method, and the analysis of the gravitational wave signal of binary black holes during the inspiral phase. We note that some of the constraints on $\alpha_{13}$ in the third column in Tab.~\ref{t-ssummary} do not perfectly match with the measurements on $\alpha_{13}$ reported in the reference in the last column (even if they are always consistent with them). This is the case in which the constraint in Tab.~\ref{t-ssummary} has been obtained by re-analyzing the data with a more recent version of our models, so the values reported in Tab.~\ref{t-ssummary} should be regarded as more accurate than those in the original paper.

{As of now, all results in the literature are consistent with the hypothesis that the spacetime around astrophysical black holes is described by the Kerr solution. As we can see from Tab.~\ref{t-ssummary}, all measurements can recover $\alpha_{13} = 0$ at 3-$\sigma$ (and in most cases, including also GRS~1716--249, a vanishing $\alpha_{13}$ is recovered even at a lower confidence level; see the original references in the last column of Tab.~\ref{t-ssummary} for the details).}
The measurement of $\alpha_{13}$ from GRS~1716--249 is quite precise, which is the result of the high photon count (the source was quite bright as it is relatively close to us), the simultaneous use of {\tt nkbb} and {\tt relxill\_nk}, and the simple spectrum of the source (there is no absorption except the Galactic one). The constraints on $\alpha_{13}$ inferred from the combination of the analysis of the thermal spectrum and the reflection features (GRS~1716--249 in this work and GRS~1915+105 and GX~339--4 in other works) are indeed among the most stringent ones. On the contrary, the sole analysis of the thermal spectrum can only provide very weak constraints, which is the case of LMC~X-1 in Tab.~\ref{t-ssummary}. The constraints on $\alpha_{13}$ from gravitational waves are not particularly stringent, but we need to note that gravitational wave tests can normally provide stringent constraints from the dynamical aspects of the theory (dipolar radiation, ringdown, inspiral-merger, consistency tests, etc.), while they are less suitable to constrain {\it ad hoc} deformations of the background metric as in this case.

{The search for possible deviations from the Kerr metric from all possible data and all possible sources with different techniques is motivated by the fact that -- within an agnostic approach like that adopted in this work -- we do not know where a possible signature of new physics would be easier to find. For example, if the no-hair theorem were violated and black holes were characterized by one more parameter in addition to the mass and the spin, the value of such a new parameter would depend on the specific object and we can expect that deviations from the Kerr metric are larger in some sources and smaller in others. In $f(R)$ gravity models, there is no uniqueness theorem and the same theory predicts the existence of different kinds of black holes~\citep{2011EPJC...71.1591S}: even if some sources are Kerr black holes, other sources may not be Kerr black holes. In similar frameworks, it is clearly mandatory to test all possible sources, because every source is potentially different. In theories with higher curvature corrections to general relativity (e.g., Einstein-dilaton-Gauss-Bonnet gravity and dynamical Chern-Simons gravity), lighter black holes present larger deviations from the Kerr metric because the curvature at the black hole horizon scales as the inverse of the square of the mass \citep{2016CQGra..33e4001Y}: in similar frameworks, it makes sense to test a specific class of objects, namely the class of sources for which we expect larger deviations from general relativity. In dynamical Chern-Simons gravity, non-rotating black holes are described by the Schwarzschild solution as in general relativity, but rotating black holes are different from the Kerr black holes, and therefore deviations from general relativity should be larger in fast-rotating objects than in slowly-rotating objects of the same mass~\citep{2009PhRvD..79h4043Y}. So the phenomenology is potentially very rich and depends on the particular gravity model.}

{Concerning the constraining power of X-ray reflection spectroscopy, it strongly depends on the specific observation and the black hole spin, while the black hole mass has only an indirect impact on the quality of the data. Most X-ray observations of accreting black holes are simply unsuitable for our tests because the spectrum has no or weak reflection features from the disk. For testing the Kerr metric, we need to select those spectra with strong reflection features, which requires that the corona illuminates well the inner part of the accretion disk. This, in turn, requires a bright corona as close as possible to the black hole. For a stellar-mass black hole transient, like GRS~1716--249, only a few observations during an outburst may be suitable for testing the Kerr metric from the analysis of the reflection features, and for some outbursts there may be no good observations at all. The inner edge of the accretion disk must be as close as possible to the black hole in order to maximize the relativistic effects in the reflection features. This, in turn, requires very fast-rotating black holes. X-ray reflection spectroscopy is unsuitable to test slowly-rotating black holes because the impact of the strong gravity region on the reflection features is too weak and, for those sources, it is normally impossible to break the degeneracy between the deformation parameter and the other parameters of the model, in particular the black hole spin, even when the quality of the data is very good. For example, this is not the case of the constraints from gravitational wave data, where the black hole spin does not play a major rule \citep{2021PhRvD.103j4036P}. Unlike all other available techniques today, X-ray reflection spectroscopy is the only one that can be used for black holes of any mass, including stellar-mass and supermassive black holes. This is because the reflection spectrum is mainly determined by atomic physics. The advantage of stellar-mass black holes over the supermassive ones is that they are normally brighter, which increases the photon count and thus decreases the Poisson noise of the source. A higher inclination angle of the accretion disk can also enhance the relativistic effects on the reflection features, so stronger constraints on the deformation parameters could be obtained from sources observed from high viewing angles. However, the factors already discussed are much more important and in Tab.~\ref{t-ssummary} we see that strong constraints on $\alpha_{13}$ can be obtained from sources with a low (e.g., GX~339--4), intermediate (e.g., GRS~1716--249), and high (e.g., EXO~1846--031 and GRS~1915+105) viewing angle.}

{Last, we note that our method is very general and can be easily applied to test any stationary and axisymmetric black hole solution with a metric known in analytic form. In the present work, we have preferred to follow an agnostic approach and employed the Johannsen metric to constrain its deformation parameter $\alpha_{13}$, but our analysis can be repeated for rotating black holes of specific modified theories of gravity. However, our current versions of {\tt nkbb} and {\tt relxill\_nk} require an analytic metric, while rotating black hole solutions beyond general relativity are often known only in numerical form. We expect to develop new versions of {\tt nkbb} and {\tt relxill\_nk} capable of working with numerical metrics in the near future in order to test specific theoretical models.}


\begin{table*}
\centering
{\renewcommand{\arraystretch}{1.3}
\begin{tabular}{lcccc}
\hline\hline
Source &  \hspace{1.0cm} Data \hspace{1.0cm}  & \hspace{0.5cm} $\alpha_{13}$ (3-$\sigma$) \hspace{0.5cm} & \hspace{0.5cm} Method \hspace{0.5cm} & \hspace{0.5cm} Main Reference \hspace{0.5cm} \\
\hline\hline
4U~1630--472 & \textsl{NuSTAR} & $-0.03_{-0.18}^{+0.63}$ & Fe-line & \citet{2021ApJ...913...79T} \\
Cygnus~X-1 & \textsl{Suzaku} & $-0.2_{-0.8}^{+0.5}$ & Fe-line & \citet{2021PhRvD.103b4055Z} \\ 
EXO~1846--031 & \textsl{NuSTAR} & $-0.03_{-0.18}^{+0.17}$ & Fe-line & \citet{2021ApJ...913...79T} \\
GRS~1716--249 & \textsl{NuSTAR}+\textsl{Swift} & $0.09_{-0.26}^{+0.02}$ & CFM + Fe-line & This work \\
GRS~1739--278 & \textsl{NuSTAR} & $-0.3_{-0.5}^{+0.6}$ & Fe-line & \citet{2021ApJ...913...79T} \\
GRS~1915+105 & \textsl{Suzaku} & $0.00_{-0.26}^{+0.17}$ & Fe-line & \citet{2019ApJ...884..147Z} \\
& \textsl{RXTE}+\textsl{Suzaku} & $0.12_{-0.27}^{+0.02}$ & CFM + Fe-line & Tripathi et al. (in preparation) \\
GS~1354--645 & \textsl{NuSTAR} & $0.0_{-0.9}^{+0.6}$ & Fe-line & \citet{2018ApJ...865..134X} \\
GW150914 & GWTC-1 & $-0.9 \pm 1.3$ & GW & \citet{2020CQGra..37m5008C} \\
GW151226 & GWTC-1 & $0.0 \pm 1.2$ & GW & \citet{2020CQGra..37m5008C} \\
GW170104 & GWTC-1 & $1.7 \pm 3.1$ & GW & \citet{2020CQGra..37m5008C} \\
GW170608 & GWTC-1 & $-0.1 \pm 0.8$ & GW & \citet{2020CQGra..37m5008C} \\
GW170814 & GWTC-1 & $-0.2 \pm 1.4$ & GW & \citet{2020CQGra..37m5008C} \\
GX~339--4 & \textsl{NuSTAR}+\textsl{Swift} & $-0.02_{-0.14}^{+0.03}$ & CFM + Fe-line & \citet{2021ApJ...907...31T} \\
LMC~X-1 & \textsl{RXTE} & $< 0.4$ & CFM & \citet{2020ApJ...897...84T} \\
Swift~J1658--4242 & \textsl{NuSTAR}+\textsl{Swift} & $0.0_{-1.0}^{+1.2}$ & Fe-line & \citet{2021ApJ...913...79T} \\
\hline\hline
\end{tabular}}
\caption{\rm Summary of the 3-$\sigma$ constraints ($\Delta\chi^2 = 9$) on the Johannsen deformation parameter $\alpha_{13}$ from stellar-mass black holes with different techniques. CFM = continuum-fitting method; Fe-line = iron-line method; GW = gravitational waves (inspiral phase). \label{t-ssummary}}
\end{table*}


\vspace{0.5cm}

{\bf Acknowledgments --}
We wish to thank Alejandro C{\'a}rdenas-Avenda{\~n}o for the gravitational wave constraints on $\alpha_{13}$ in Tab.~\ref{t-ssummary}.
This work was supported by the Innovation Program of the Shanghai Municipal Education Commission, Grant No.~2019-01-07-00-07-E00035, the National Natural Science Foundation of China (NSFC), Grant No.~11973019, and Fudan University, Grant No.~JIH1512604. 
D.A. is supported through the Teach@T{\"u}bingen Fellowship.


\appendix

\section{Johannsen metric}

For the convenience of the readers, we report here the expression of the Johannsen metric~\citep{2013PhRvD..88d4002J}. In Boyer-Lindquist-like coordinates, the line element reads
\be\label{eq-jm}
ds^2 &=&-\frac{\tilde{\Sigma}\left(\Delta-a^2A_2^2\sin^2\theta\right)}{B^2}dt^2 
+\frac{\tilde{\Sigma}}{\Delta A_5}dr^2+\tilde{\Sigma} d\theta^2 \nonumber\\
&& -\frac{2a\left[\left(r^2+a^2\right)A_1A_2-\Delta\right]\tilde{\Sigma}\sin^2\theta}{B^2}dtd\phi 
+\frac{\left[\left(r^2+a^2\right)^2A_1^2-a^2\Delta\sin^2\theta\right]\tilde{\Sigma}\sin^2\theta}{B^2}d\phi^2 \, ,
\ee
where $M$ is the black hole mass, $a = J/M$, $J$ is the black hole spin angular momentum, $\tilde{\Sigma} = \Sigma + f$, and
\be
\Sigma = r^2 + a^2 \cos^2\theta \, , \qquad
\Delta = r^2 - 2 M r + a^2 \, , \qquad
B = \left(r^2+a^2\right)A_1-a^2A_2\sin^2\theta \, .
\ee
The functions $f$, $A_1$, $A_2$, and $A_5$ are defined as
\be\label{eq-fa1a2a5}
f = \sum^\infty_{n=3} \epsilon_n \frac{M^n}{r^{n-2}} \, , \quad
A_1 = 1 + \sum^\infty_{n=3} \alpha_{1n} \left(\frac{M}{r}\right)^n \, , \quad
A_2 = 1 + \sum^\infty_{n=2} \alpha_{2n}\left(\frac{M}{r}\right)^n \, , \quad
A_5 = 1 + \sum^\infty_{n=2} \alpha_{5n}\left(\frac{M}{r}\right)^n \, ,
\ee
where $\{ \epsilon_n \}$, $\{ \alpha_{1n} \}$, $\{ \alpha_{2n} \}$, and $\{ \alpha_{5n} \}$ are four infinite sets of deformation parameters without constraints from the Newtonian limit and Solar System experiments.

The Johannsen metric is not a solution of any particular theory of gravity. It was proposed in \citet{2013PhRvD..88d4002J} for agnostic tests of the Kerr metric with electromagnetic data. The metric is obtained by imposing that the spacetime is regular outside of the event horizon (no naked singularities, closed time-like curves, etc.) and has a Carter-like constant (i.e., the equations of motion of a test-particle can be written in first-order form). Under these conditions, one can find that the resulting metric is characterized by four free functions ($f$, $A_1$, $A_2$, and $A_5$).

In the present paper, for the sake of simplicity, we have only considered the deformation parameter $\alpha_{13}$ and all other deformation parameters are assumed to vanish. In order to work with a regular metric, we need to impose the following constraints on the black hole spin parameter $a_*$ and the deformation parameter $\alpha_{13}$
\be\label{eq-app-c}
- 1 \le a_* \le 1 \, , \quad \alpha_{13} > - \frac{1}{2} \left( 1 + \sqrt{1 - a^2_*} \right)^4 \, .
\ee



\begin{thebibliography}{99}

\bibitem[Abbott et al.(2016)]{2016PhRvL.116v1101A} Abbott, B.~P., Abbott, R., Abbott, T.~D., et al.\ 2016, \prl, 116, 221101. doi:10.1103/PhysRevLett.116.221101

\bibitem[Abbott et al.(2019a)]{2019PhRvD.100j4036A} Abbott, B.~P., Abbott, R., Abbott, T.~D., et al.\ 2019a, \prd, 100, 104036. doi:10.1103/PhysRevD.100.104036

\bibitem[Abdikamalov et al.(2019b)]{2019ApJ...878...91A} Abdikamalov, A.~B., Ayzenberg, D., Bambi, C., et al.\ 2019b, \apj, 878, 91. doi:10.3847/1538-4357/ab1f89

\bibitem[Abdikamalov et al.(2020)]{2020ApJ...899...80A} Abdikamalov, A.~B., Ayzenberg, D., Bambi, C., et al.\ 2020, \apj, 899, 80. doi:10.3847/1538-4357/aba625

\bibitem[Ballet et al.(1993)]{1993IAUC.5874....1B} Ballet, J., Denis, M., Gilfanov, M., et al.\ 1993, \iaucirc, 5874

\bibitem[Bambi(2013)]{2013PhRvD..87b3007B} Bambi, C.\ 2013, \prd, 87, 023007. doi:10.1103/PhysRevD.87.023007

\bibitem[Bambi(2017a)]{2017RvMP...89b5001B} Bambi, C.\ 2017a, Reviews of Modern Physics, 89, 025001. doi:10.1103/RevModPhys.89.025001

\bibitem[Bambi(2017b)]{2017bhlt.book.....B} Bambi, C.\ 2017b, Black Holes: A Laboratory for Testing Strong Gravity, ISBN 978-981-10-4523-3. Springer Nature Singapore Pte Ltd., 2017. doi:10.1007/978-981-10-4524-0

\bibitem[Bambi \& Barausse(2011)]{2011ApJ...731..121B} Bambi, C. \& Barausse, E.\ 2011, \apj, 731, 121. doi:10.1088/0004-637X/731/2/121

\bibitem[Bambi et al.(2014)]{2014PhRvD..89l7302B} Bambi, C., Malafarina, D., \& Tsukamoto, N.\ 2014, \prd, 89, 127302. doi:10.1103/PhysRevD.89.127302

\bibitem[Bambi et al.(2017)]{2017ApJ...842...76B} Bambi, C., C{\'a}rdenas-Avenda{\~n}o, A., Dauser, T., et al.\ 2017, \apj, 842, 76. doi:10.3847/1538-4357/aa74c0

\bibitem[Bambi et al.(2019)]{2019PhRvD.100d4057B} Bambi, C., Freese, K., Vagnozzi, S., et al.\ 2019, \prd, 100, 044057. doi:10.1103/PhysRevD.100.044057

\bibitem[Bambi et al.(2021)]{2021SSRv..217...65B} Bambi, C., Brenneman, L.~W., Dauser, T., et al.\ 2021, \ssr, 217, 65. doi:10.1007/s11214-021-00841-8

\bibitem[Bassi et al.(2019)]{2019MNRAS.482.1587B} Bassi, T., Del Santo, M., D'A{\i}, A., et al.\ 2019, \mnras, 482, 1587. doi:10.1093/mnras/sty2739
  
\bibitem[Brenneman \& Reynolds(2006)]{2006ApJ...652.1028B} Brenneman, L.~W. \& Reynolds, C.~S.\ 2006, \apj, 652, 1028. doi:10.1086/508146  
  
\bibitem[Cao et al.(2018)]{2018PhRvL.120e1101C} Cao, Z., Nampalliwar, S., Bambi, C., et al.\ 2018, \prl, 120, 051101. doi:10.1103/PhysRevLett.120.051101   
  
\bibitem[C{\'a}rdenas-Avenda{\~n}o et al.(2020)]{2020CQGra..37m5008C} C{\'a}rdenas-Avenda{\~n}o, A., Nampalliwar, S., \& Yunes, N.\ 2020, Classical and Quantum Gravity, 37, 135008. doi:10.1088/1361-6382/ab8f64  
  
\bibitem[Cardoso \& Pani(2019)]{2019LRR....22....4C} Cardoso, V. \& Pani, P.\ 2019, Living Reviews in Relativity, 22, 4. doi:10.1007/s41114-019-0020-4  

\bibitem[Chatterjee et al.(2021)]{2021Ap&SS.366...63C} Chatterjee, K., Debnath, D., Chatterjee, D., et al.\ 2021, \apss, 366, 63. doi:10.1007/s10509-021-03967-x
  
\bibitem[Chru{\'s}ciel et al.(2012)]{2012LRR....15....7C} Chru{\'s}ciel, P.~T., Costa, J.~L., \& Heusler, M.\ 2012, Living Reviews in Relativity, 15, 7. doi:10.12942/lrr-2012-7  
 
\bibitem[Dauser et al.(2010)]{2010MNRAS.409.1534D} Dauser, T., Wilms, J., Reynolds, C.~S., et al.\ 2010, \mnras, 409, 1534. doi:10.1111/j.1365-2966.2010.17393.x
  
\bibitem[Dauser et al.(2013)]{2013MNRAS.430.1694D} Dauser, T., Garcia, J., Wilms, J., et al.\ 2013, \mnras, 430, 1694. doi:10.1093/mnras/sts710  
  
\bibitem[Debnath et al.(2010)]{2010A&A...520A..98D} Debnath, D., Chakrabarti, S.~K., \& Nandi, A.\ 2010, \aap, 520, A98. doi:10.1051/0004-6361/201014990  
  
\bibitem[della Valle et al.(1994)]{1994A&A...290..803D} della Valle, M., Mirabel, I.~F., \& Rodriguez, L.~F.\ 1994, \aap, 290, 803  
  
\bibitem[Fabian et al.(1989)]{1989MNRAS.238..729F} Fabian, A.~C., Rees, M.~J., Stella, L., et al.\ 1989, \mnras, 238, 729. doi:10.1093/mnras/238.3.729  
  
\bibitem[Gammie(1999)]{1999ApJ...522L..57G} Gammie, C.~F.\ 1999, \apjl, 522, L57. doi:10.1086/312207  
  
\bibitem[Garc{\'\i}a et al.(2014)]{2014ApJ...782...76G} Garc{\'\i}a, J., Dauser, T., Lohfink, A., et al.\ 2014, \apj, 782, 76. doi:10.1088/0004-637X/782/2/76
  
\bibitem[Giddings(2017)]{2017NatAs...1E..67G} Giddings, S.~B.\ 2017, Nature Astronomy, 1, 0067. doi:10.1038/s41550-017-0067  
  
\bibitem[Gonzalez et al.(2017)]{2017MNRAS.472.1932G} Gonzalez, A.~G., Wilkins, D.~R., \& Gallo, L.~C.\ 2017, \mnras, 472, 1932. doi:10.1093/mnras/stx2080  
  
\bibitem[Harmon et al.(1994)]{1994IAUC.6104....1H} Harmon, B.~A., Zhang, S.~N., Paciesas, W.~S., et al.\ 1994, \iaucirc, 6104  
  
\bibitem[Herdeiro \& Radu(2014)]{2014PhRvL.112v1101H} Herdeiro, C.~A.~R. \& Radu, E.\ 2014, \prl, 112, 221101. doi:10.1103/PhysRevLett.112.221101  
  
\bibitem[Johannsen(2013)]{2013PhRvD..88d4002J} Johannsen, T.\ 2013, \prd, 88, 044002. doi:10.1103/PhysRevD.88.044002  
  
\bibitem[Johannsen \& Psaltis(2013)]{2013ApJ...773...57J} Johannsen, T. \& Psaltis, D.\ 2013, \apj, 773, 57. doi:10.1088/0004-637X/773/1/57  
  
\bibitem[Jiang et al.(2020)]{2020MNRAS.492.1947J} Jiang, J., F{\"u}rst, F., Walton, D.~J., et al.\ 2020, \mnras, 492, 1947. doi:10.1093/mnras/staa017  
  
\bibitem[Kerr(1963)]{1963PhRvL..11..237K} Kerr, R.~P.\ 1963, \prl, 11, 237. doi:10.1103/PhysRevLett.11.237  
  
\bibitem[Kleihaus et al.(2011)]{2011PhRvL.106o1104K} Kleihaus, B., Kunz, J., \& Radu, E.\ 2011, \prl, 106, 151104. doi:10.1103/PhysRevLett.106.151104  

\bibitem[Krolik(1999)]{1999ApJ...515L..73K} Krolik, J.~H.\ 1999, \apjl, 515, L73. doi:10.1086/311979

\bibitem[Kulkarni et al.(2011)]{2011MNRAS.414.1183K} Kulkarni, A.~K., Penna, R.~F., Shcherbakov, R.~V., et al.\ 2011, \mnras, 414, 1183. doi:10.1111/j.1365-2966.2011.18446.x

\bibitem[Lu \& Torres(2003)]{2003IJMPD..12...63L} Lu, Y. \& Torres, D.~F.\ 2003, International Journal of Modern Physics D, 12, 63. doi:10.1142/S0218271803002718
  
\bibitem[Masetti et al.(1996)]{1996A&A...314..123M} Masetti, N., Bianchini, A., Bonibaker, J., et al.\ 1996, \aap, 314, 123  

\bibitem[Miniutti et al.(2003)]{2003MNRAS.344L..22M} Miniutti, G., Fabian, A.~C., Goyder, R., et al.\ 2003, \mnras, 344, L22. doi:10.1046/j.1365-8711.2003.06988.x
  
\bibitem[McClintock et al.(2006)]{2006ApJ...652..518M} McClintock, J.~E., Shafee, R., Narayan, R., et al.\ 2006, \apj, 652, 518. doi:10.1086/508457  
  
\bibitem[McClintock et al.(2014)]{2014SSRv..183..295M} McClintock, J.~E., Narayan, R., \& Steiner, J.~F.\ 2014, \ssr, 183, 295. doi:10.1007/s11214-013-0003-9    
  
\bibitem[Noble et al.(2010)]{2010ApJ...711..959N} Noble, S.~C., Krolik, J.~H., \& Hawley, J.~F.\ 2010, \apj, 711, 959. doi:10.1088/0004-637X/711/2/959  
  
\bibitem[Novikov \& Thorne(1973)]{1973blho.conf..343N} Novikov, I.~D. \& Thorne, K.~S.\ 1973, Black Holes (Les Astres Occlus), 343  
  
\bibitem[Page \& Thorne(1974)]{1974ApJ...191..499P} Page, D.~N. \& Thorne, K.~S.\ 1974, \apj, 191, 499. doi:10.1086/152990  
  
\bibitem[Penna et al.(2010)]{2010MNRAS.408..752P} Penna, R.~F., McKinney, J.~C., Narayan, R., et al.\ 2010, \mnras, 408, 752. doi:10.1111/j.1365-2966.2010.17170.x  
  
\bibitem[Penna et al.(2012)]{2012MNRAS.420..684P} Penna, R.~F., Sadowski, A., \& McKinney, J.~C.\ 2012, \mnras, 420, 684. doi:10.1111/j.1365-2966.2011.20084.x  
  
\bibitem[Psaltis et al.(2020)]{2020PhRvL.125n1104P} Psaltis, D., Medeiros, L., Christian, P., et al.\ 2020, \prl, 125, 141104. doi:10.1103/PhysRevLett.125.141104  

\bibitem[Psaltis et al.(2021)]{2021PhRvD.103j4036P} Psaltis, D., Talbot, C., Payne, E., et al.\ 2021, \prd, 103, 104036. doi:10.1103/PhysRevD.103.104036
  
\bibitem[Pun et al.(2008)]{2008PhRvD..78b4043P} Pun, C.~S.~J., Kov{\'a}cs, Z., \& Harko, T.\ 2008, \prd, 78, 024043. doi:10.1103/PhysRevD.78.024043  
  
\bibitem[Riaz et al.(2020)]{2020arXiv201207469R} Riaz, S., Abdikamalov, A.~B., Ayzenberg, D., et al.\ 2020, arXiv:2012.07469  

\bibitem[Roy et al.(2021)]{2021PhRvD.104d4001R} Roy, R., Abdikamalov, A.~B., Ayzenberg, D., et al.\ 2021, \prd, 104, 044001. doi:10.1103/PhysRevD.104.044001
  
\bibitem[Schee \& Stuchl{\'\i}k(2009)]{2009GReGr..41.1795S} Schee, J. \& Stuchl{\'\i}k, Z.\ 2009, General Relativity and Gravitation, 41, 1795. doi:10.1007/s10714-008-0753-y  
  
\bibitem[Sebastiani \& Zerbini(2011)]{2011EPJC...71.1591S} Sebastiani, L. \& Zerbini, S.\ 2011, European Physical Journal C, 71, 1591. doi:10.1140/epjc/s10052-011-1591-8  
  
\bibitem[Tao et al.(2019)]{2019ApJ...887..184T} Tao, L., Tomsick, J.~A., Qu, J., et al.\ 2019, \apj, 887, 184. doi:10.3847/1538-4357/ab5282  
  
\bibitem[Torres(2002)]{2002NuPhB.626..377T} Torres, D.~F.\ 2002, Nuclear Physics B, 626, 377. doi:10.1016/S0550-3213(02)00038-X    
  
\bibitem[Tripathi et al.(2019)]{2019ApJ...875...56T} Tripathi, A., Nampalliwar, S., Abdikamalov, A.~B., et al.\ 2019, \apj, 875, 56. doi:10.3847/1538-4357/ab0e7e 

\bibitem[Tripathi et al.(2020)]{2020ApJ...897...84T} Tripathi, A., Zhou, M., Abdikamalov, A.~B., et al.\ 2020, \apj, 897, 84. doi:10.3847/1538-4357/ab9600 

\bibitem[Tripathi et al.(2021a)]{2021ApJ...907...31T} Tripathi, A., Abdikamalov, A.~B., Ayzenberg, D., et al.\ 2021a, \apj, 907, 31. doi:10.3847/1538-4357/abccbd

\bibitem[Tripathi et al.(2021b)]{2021ApJ...913...79T} Tripathi, A., Zhang, Y., Abdikamalov, A.~B., et al.\ 2021b, \apj, 913, 79. doi:10.3847/1538-4357/abf6cd

\bibitem[Tripathi et al.(2021c)]{2021JCAP...07..002T} Tripathi, A., Zhou, B., Abdikamalov, A.~B., et al.\ 2021c, \jcap, 2021, 002. doi:10.1088/1475-7516/2021/07/002
 
\bibitem[Vagnozzi \& Visinelli(2019)]{2019PhRvD.100b4020V} Vagnozzi, S. \& Visinelli, L.\ 2019, \prd, 100, 024020. doi:10.1103/PhysRevD.100.024020  
  
\bibitem[Wilkins \& Fabian(2012)]{2012MNRAS.424.1284W} Wilkins, D.~R. \& Fabian, A.~C.\ 2012, \mnras, 424, 1284. doi:10.1111/j.1365-2966.2012.21308.x  
  
\bibitem[Will(2014)]{2014LRR....17....4W} Will, C.~M.\ 2014, Living Reviews in Relativity, 17, 4. doi:10.12942/lrr-2014-4  
  
\bibitem[Wilms et al.(2000)]{2000ApJ...542..914W} Wilms, J., Allen, A., \& McCray, R.\ 2000, \apj, 542, 914. doi:10.1086/317016  

\bibitem[Xu et al.(2018)]{2018ApJ...865..134X} Xu, Y., Nampalliwar, S., Abdikamalov, A.~B., et al.\ 2018, \apj, 865, 134. doi:10.3847/1538-4357/aadb9d
  
\bibitem[Yagi \& Stein(2016)]{2016CQGra..33e4001Y} Yagi, K. \& Stein, L.~C.\ 2016, Classical and Quantum Gravity, 33, 054001. doi:10.1088/0264-9381/33/5/054001  

\bibitem[Yunes \& Pretorius(2009)]{2009PhRvD..79h4043Y} Yunes, N. \& Pretorius, F.\ 2009, \prd, 79, 084043. doi:10.1103/PhysRevD.79.084043  
  
\bibitem[Yunes et al.(2016)]{2016PhRvD..94h4002Y} Yunes, N., Yagi, K., \& Pretorius, F.\ 2016, \prd, 94, 084002. doi:10.1103/PhysRevD.94.084002  
  
\bibitem[Zhang et al.(1997)]{1997ApJ...482L.155Z} Zhang, S.~N., Cui, W., \& Chen, W.\ 1997, \apjl, 482, L155. doi:10.1086/310705  
  
\bibitem[Zhang et al.(2019)]{2019ApJ...884..147Z} Zhang, Y., Abdikamalov, A.~B., Ayzenberg, D., et al.\ 2019, \apj, 884, 147. doi:10.3847/1538-4357/ab4271  
  
\bibitem[Zhang et al.(2021)]{2021PhRvD.103b4055Z} Zhang, Z., Liu, H., Abdikamalov, A.~B., et al.\ 2021, \prd, 103, 024055. doi:10.1103/PhysRevD.103.024055  
  
\bibitem[Zhou et al.(2021)]{2021JCAP...01..047Z} Zhou, B., Abdikamalov, A.~B., Ayzenberg, D., et al.\ 2021, \jcap, 2021, 047. doi:10.1088/1475-7516/2021/01/047  
  
\bibitem[Zhou et al.(2019)]{2019PhRvD..99j4031Z} Zhou, M., Abdikamalov, A.~B., Ayzenberg, D., et al.\ 2019, \prd, 99, 104031. doi:10.1103/PhysRevD.99.104031  

\bibitem[Zhu et al.(2020)]{2020EPJC...80..622Z} Zhu, J., Abdikamalov, A.~B., Ayzenberg, D., et al.\ 2020, European Physical Journal C, 80, 622. doi:10.1140/epjc/s10052-020-8198-x

\bibitem[Zhu et al.(2012)]{2012MNRAS.424.2504Z} Zhu, Y., Davis, S.~W., Narayan, R., et al.\ 2012, \mnras, 424, 2504. doi:10.1111/j.1365-2966.2012.21181.x
  
\end{thebibliography}
\end{document}